\documentclass[11pt,a4paper]{article}
\pdfoutput=1 

\usepackage{jheppub}
\usepackage{amssymb,amsmath,amsthm,graphicx}
\usepackage{graphics, color}
\usepackage{subfigure}
\usepackage{latexsym}
\usepackage{bm}
\usepackage{epsfig}
\usepackage{textcomp}
\usepackage[utf8]{inputenc}
\allowdisplaybreaks[1]
\def \dd {\mathrm{d}}
\def\ie{{\it i.e.} }

\begin{document}
\title{Gregory-Laflamme and Superradiance encounter Black Resonator Strings}
\author[1]{\'Oscar J.~C.~Dias,}
\author[2]{Takaaki~Ishii,}
\author[3]{Keiju~Murata,}
\author[4]{Jorge~E.~Santos,}
\author[5]{Benson~Way}

\affiliation[1]{STAG research centre and Mathematical Sciences, University of Southampton, University Road, Southampton SO17 1BJ, UK.}
\affiliation[2]{Department of Physics, Rikkyo University, Nishi-Ikebukuro, Tokyo 171-8501, Japan}
\affiliation[3]{Department of Physics, College of Humanities and Sciences, Nihon University, Sakurajosui, Tokyo 156-8550, Japan}
\affiliation[4]{DAMTP, Centre for Mathematical Sciences, University of Cambridge, Wilberforce Road, Cambridge CB3 0WA, UK}
\affiliation[5]{Departament de F\'{i}sica Qu\`{a}ntica i Astrof\'{i}sica, Institut de Ci\`{e}ncies del Cosmos\\
Universitat de Barcelona, Mart\'{i} i Franqu\`{e}s, 1, E-08028 Barcelona, Spain}
\emailAdd{ojcd1r13@soton.ac.uk}
\emailAdd{ishiitk@rikkyo.ac.jp}
\emailAdd{murata.keiju@nihon-u.ac.jp}
\emailAdd{jss55@cam.ac.uk}
\emailAdd{benson@icc.ub.edu}

\newcommand{\blue}{\color{blue}}
\newcommand{\red}{\color{red}}

\abstract{
We construct novel black strings that are neither time-translation invariant, nor axisymmetric, nor translationally invariant in the string direction, but nevertheless have a helical Killing vector field. These solutions branch from the superradiant instability of $D=6$ Myers-Perry black strings with equal angular momenta. We coin these solutions as {\it black resonator strings} and we find that they have more entropy than Myers-Perry black strings for the energies and angular momenta where both solutions coexist.  We also construct Kaluza-Klein geons, which share the symmetries of black resonator strings, but are horizonless.  Unlike in other superradiant systems, Kaluza-Klein geons are not the horizonless limit of black resonator strings and are instead entirely separate solutions.
}
\preprint{RUP-22-26}
\maketitle

\section{Introduction}
Over the past three decades, the (Schwarzschild) black strings and black branes \cite{Horowitz:1991cd} have served as the model solution for studying the physics of black holes in higher dimensions.  They produce a plethora of physical phenomena not found in four-dimensional gravity, including the Gregory-Laflamme instability \cite{Gregory:1993vy,Gregory:1994bj}, non-uniqueness \cite{Gubser:2001ac,Harmark:2002tr,Kol:2002xz,Wiseman:2002zc,Kol:2003ja,Harmark:2003dg,Harmark:2003yz,Kudoh:2003ki,Sorkin:2004qq,Gorbonos:2004uc,Kudoh:2004hs,Dias:2007hg,Harmark:2007md,Wiseman:2011by,Figueras:2012xj,Dias:2017coo}, and violations of the weak cosmic censorship \cite{Lehner:2010pn,Figueras:2022zkg}.  The same physics is found in other higher dimensional black holes, including black rings \cite{Emparan:2001wn,Santos:2015iua,Figueras:2015hkb}, ultraspinning black holes \cite{Dias:2009iu,Dias:2010maa,Dias:2010eu,Dias:2010maa,Dias:2010gk,Dias:2011jg,Emparan:2003sy,Dias:2014cia,Emparan:2014pra,Dias:2015nua}, and certain black holes in supergravity \cite{Banks:1998dd,Peet:1998cr,Hubeny:2002xn,Dias:2015pda,Buchel:2015gxa,Dias:2016eto,Cardona:2020unx}.

Recently in \cite{Dias:2022mde}, we studied the $D=6$ equal-spinning Myers-Perry black string.  There, we found a different instability that affects certain rotating black strings: the superradiant instability \cite{Marolf:2004fya,Cardoso:2004zz,Dias:2006zv}.  In this instability, the ergoregion of the black string amplifies incoming waves.  Meanwhile, momentum along the string direction produces an effective mass term, allowing these waves to be confined.  The amplification process then continues until an instability takes over.

In other superradiant instabilities (say, in AdS), there are new solutions that branch from its onset called black resonators \cite{Dias:2015rxy,Ishii:2018oms,Ishii:2020muv,Ishii:2021xmn}.  The same is true for rotating black strings, and by analogy, we call these \emph{black resonator strings}. Linear evidence for the existence of black resonator strings was identified in \cite{Dias:2022mde}, but here we construct them explicitly and study their properties.  We will do so both by using higher-order perturbation theory and by solving the full Einstein equation numerically.  As in \cite{Dias:2022mde}, we focus on the case where the $D=6$ equal-spinning Myers-Perry black string is a solution. The mechanism for the instability, as well as the prediction for the existence of novel solutions branching from its onset, has been anticipated in \cite{Marolf:2004fya}\footnote{Indeed, our black string resonators were coined `gyratons' in \cite{Marolf:2004fya}.}.

We will find some similarities between black resonator strings and black resonators. Both solutions are rotating black objects with gravitational hair.  Both are neither time-translation invariant nor axisymmetric, but they are nevertheless time-periodic\footnote{These two statements together imply that black resonator strings have a  helical Killing vector field $K \equiv \partial_t+\Omega_H \partial_\psi$, where $\partial_t$ and $\partial_\psi$ are the asymptotic Killing vector fields generating time translations and translations in the angular direction of the equal-spin rotation,
respectively.}. Both have a higher entropy than the unstable solution from which they branched off (for fixed conserved charges), \ie it is permissible, entropically, for the superradiant instability to evolve towards these solutions. 

However, we also find some differences.  Unlike black resonators, black resonator strings are expected to be stable to perturbations with arbitrarily high wavenumber. 
Unlike black resonators, where their horizonless limit is a nontrivial solution called a geon, black resonator strings do not have a zero-horizon limit.  Curiously, we will also construct Kaluza-Klein geons, which are horizonless configurations that share the same symmetries as black resonator strings, but these appear to be entirely disconnected and seemingly unrelated to black resonator strings.

The plan of the manuscript is as follows. Section \ref{sec:MPinstabilities} is dedicated to review the equal angular momenta Myers-Perry black string and the properties of its Gregory-Laflamme and superradiant instabilities studied in \cite{Dias:2022mde}. In section~\ref{sec:ansatzThermo} we motivate the ansatz we use to find the black resonator strings as well as their thermodynamic quantities. In section~\ref{sec:perturbative}, we provide the details to our perturbative calculation. We describe our construction of Kaluza-Klein geons in section~\ref{sec:geons}, then explain in section~\ref{sec:warmholes} why the theory does not have Kaluza-Klein warm holes (of the type found in \cite{Dias:2021vve}).  Finally, we piece together the phase diagram of solutions in section~\ref{sec:PhaseDiag} and end with a discussion in section~\ref{sec:Discussion} where we speculate on evolution scenarios.

\section{Myers-Perry string and its instabilities}\label{sec:MPinstabilities}

The Myers-Perry (MP) black hole \cite{Myers:1986un} is a rotating higher-dimensional black hole that is asymptotically flat and Ricci flat ($R_{\mu\nu}=0$).  In five dimensions, the Myers-Perry black hole has three dimensionful parameters, a common choice being the mass radius parameter $r_0$ and two angular momentum parameters $a_1$ and $a_2$ \cite{Myers:1986un,Hawking_1999}. In general, this solution has an isometry group $\mathbb{R}_t \times U(1)^2$, but in the equal-spinning case $a_1=a_2 \equiv a$, the solution has an enhanced symmetry $\mathbb{R}_t \times U(2)$ \cite{Gibbons:2004js,Gibbons_2005}.  The extra symmetries simplify the study of rotating black holes.

We are interested in the associated 6-dimensional rotating black string, which is the product spacetime of an equal-spinning Myers-Perry black hole with a circle. Hereafter, we simply refer to this solution as the Myers-Perry black string. Its metric can be written:\footnote{The radial coordinate used here can be related to the standard Boyer-Lindquist radial coordinate of \cite{Myers:1986un} through $r^2 \to r^2+a^2$. We shall use a notation where capital Latin indices run over the 6-dimensional coordinates, lower case Latin indices run over all spatial coordinates except the radial one, and Greek indices run over all coordinates except the extended/string direction.}
\begin{eqnarray} \label{MPstring}
 && {\mathrm d}s^2 _{\rm MP \,string}= -\frac{F}{H}\,{\mathrm d}t^2 +\frac{{\mathrm d}r^2}{F} + r^2 \left[ H\left( \frac{\sigma_3}{2} -\frac{\Omega}{H}\, {\mathrm d}t \right)^2+ {\mathrm d}s^2_{\mathbb{C}\mathrm{P}^1}  \right] + {\rm d}z^2\,,
\end{eqnarray}
where
\begin{equation}
{\mathrm d}s^2_{\mathbb{C}\mathrm{P}^1}=
\frac{1}{4}\left( \sigma_1^2 +\sigma_2^2\right)
\end{equation}
is the metric of the complex projective space $\mathbb{C}\mathrm{P}^1$, which is isomorphic to the 2-sphere $S^2$\footnote{In odd spacetime dimensions, $D=2N+3$, equal-spinning Myers-Perry black holes have a homogeneously squashed $S^{2N+1}$ written as an $S^1$ fibred over $\mathbb{C}\mathrm{P}^{N}$.  Note that the Fubini-Study line element ${\mathrm d}s^2_{\mathbb{C}\mathrm{P}^1}$ on $\mathbb{C}\mathrm{P}^{1}$ is simply the familiar metric for the $S^2$ and the  K\"ahler potential $A$ of $\mathbb{C}\mathrm{P}^1$ is simply $A=\frac{1}{2}\cos\theta$. For further details see  \cite{Gibbons:2004js,Gibbons_2005,Kunduri:2006qa,Dias:2010eu}.}.
Letting $(\theta,\phi,\psi)$ be the Euler angles of a $S^3$ with ranges $0\leq \theta \leq \pi $, $0\leq \phi <2\pi$, and $0\leq \psi <2\pi$, we have defined the 1-forms $\sigma_{\imath}$ ($\imath=1,2,3$) on $S^3$ as
\begin{equation}\label{Eulerforms}
\begin{split}
  \sigma_1 &= -\sin(2 \psi) \, \mathrm{d}\theta + \cos(2 \psi)\sin\theta \,\mathrm{d}\phi\ ,\\
  \sigma_2 &= \cos(2 \psi) \,\mathrm{d}\theta + \sin(2\psi)\sin\theta \,\mathrm{d}\phi\ ,\\
  \sigma_3 &= 2\,\mathrm{d}\psi + \cos\theta \,\mathrm{d}\phi  \,.
\end{split}
\end{equation}
which satisfy the Maurer-Cartan equation $\mathrm{d}\sigma_\imath = \frac{1}{2} \epsilon_{\imath \jmath k} \sigma_\jmath \wedge \sigma_k$. In \eqref{MPstring}, we have also defined
\begin{equation}\label{MPfns}
F(r)= 1-\frac{r_0^2}{r^2}+\frac{a^2r_0^2}{r^4}\,, \qquad H(r)=1+ \frac{a^2 r_0^2}{r^4}\,, \qquad
\Omega=\frac{a \,r_0^2}{r^4}\,.
\end{equation}
This solution has an event horizon at $r=r_+$  (the largest real root of $F$) with Killing horizon generator $K=\partial_t+\Omega_H \partial_\psi$, where $\Omega_H\equiv \Omega(r_+)/H(r_+)$ is the horizon angular velocity. We can express the mass radius parameter as $r_0=r_+^2/\sqrt{r_+^2-a^2}$.

Letting $L$ be the length of the extended direction $z$ (so $z\sim z+L$) and introducing the dimensionless rotation parameter $\widetilde{a} \equiv a/r_+$, the energy, angular momentum, tension along $z$, temperature,  entropy and angular velocity of the Myers-Perry black string are, respectively:
\begin{equation}\label{ThermoMP}
\begin{split}
  E\big|_{\hbox{\tiny MP}} &= \frac{L}{8 G_6} \frac{3 \pi  r_+^2}{1-\widetilde{a} ^2} \,, \qquad J\big|_{\hbox{\tiny MP}} =  \frac{L}{4 G_6} \frac{\pi  \widetilde{a}  r_+^3}{1-\widetilde{a} ^2} \,, \qquad T_z\big|_{\hbox{\tiny MP}}= \frac{1}{8 G_6 } \frac{\pi  r_+^2}{1-\widetilde{a} ^2}  \,,  \\
  T_H\big|_{\hbox{\tiny MP}} &= \frac{1}{2 \pi  r_+} \frac{1-2 \widetilde{a} ^2}{\sqrt{1-\widetilde{a} ^2}}\,, \qquad S_{H}\big|_{\hbox{\tiny MP}} = \frac{L}{G_6} \frac{\pi ^2 r_+^3}{2 \sqrt{1-\widetilde{a} ^2}}\,, \qquad \Omega_H \big|_{\hbox{\tiny MP}} = \frac{\widetilde{a} }{r_+} \,. \\
\end{split}
\end{equation}
At $\widetilde{a}=1/\sqrt{2}$, the temperature vanishes at extremality.  Note that we decided to measure $J$ and $\Omega_H$ with respect to $\psi\sim \psi+2\pi$, in accordance with the work of Myers and Perry\footnote{\label{footCanPsi}In these conventions the superradiant factor (also known as resonant or syncronization factor), that will appear later in our discussions, reads $\omega-2m\Omega_H$. The non-standard factor of 2 reflects the fact that we have equal rotation along the two planes and we are using the period of $2\pi$ for $\psi$.} \cite{Myers:1986un}.

The Myers-Perry black string \eqref{MPstring} is unstable to (at least) two sectors of perturbations \cite{Dias:2022mde}. One is the familiar Gregory-Laflamme instability, where horizons with a large separation of length scales are unstable to forming ripples along the extended directions. This instability was first studied in the context of Schwarzschild black strings/branes  \cite{Gregory:1993vy,Gregory:1994bj}, later extended to rotating black strings in \cite{Dias:2010eu,Dias:2009iu,Dias:2010maa,Dias:2011jg}, and then studied in detail for the Myers-Perry black string in \cite{Dias:2022mde}.

The other instability involves superradiant scattering, where an ergoregion of a rotating black hole amplifies incident waves.  If these waves are reflected or confined, the amplification process continues, leading to an exponentially growing instability. In black strings, the Kaluza-Klein momentum along the string direction can provide an effective mass term that confines superradiant bound states \cite{Marolf:2004fya}. In the Myers-Perry black string~\eqref{MPstring}, this superradiant instability was found and studied in \cite{Dias:2022mde} (see also \cite{Cardoso:2004zz,Dias:2006zv} for perturbative analysis of this instability in Kerr strings for scalar fields).

The dominant superradiant instability of \eqref{MPstring} is expected to be sourced by the  perturbation of the form $\mathrm{d}s^2= \mathrm{d}s^2_\mathrm{\rm MP \,string}+ \delta \mathrm{d}s^2_{\hbox{\tiny SR}} $ with  \cite{Dias:2022mde}
\begin{equation}\label{SRpert}
\delta \mathrm{d}s^2_{\hbox{\tiny SR}}= \frac{1}{4}e^{{\rm i} k z} e^{-{\rm i}\omega t} e^{{\rm i}\,(m-2)\,(\phi+2\psi)} \cos^{2(m-2)}\left(\frac{\theta}{2}\right) \,Q(r)\left(\sigma_1^2-\sigma_2^2+2\,{\rm i}\,\sigma_1\,\sigma_2\right),
\end{equation}
which describes a superradiant perturbation with a charged scalar harmonic of $\mathbb{C}\mathrm{P}^{1}$.  Here, $k=2\pi/L$, and $m$ is a half-integer ($m=2,5/2,3,7/2, 4,\cdots$) corresponding to the azimuthal quantum number for $\psi$ \cite{Hartnett:2013fba,Ishii:2020muv}.  We are most interested in the $m=2$ case as the instability of this perturbation has the highest growth rate and it does not depend on the polar angle~$\theta$.

It is often convenient for us to work in units of the horizon radius $r_+$.  Accordingly, we introduce the dimensionless quantities
\begin{equation}\label{dimAux}
\widetilde{a}=a/r_+\,, \quad \widetilde{\Omega}_H=\Omega_H r_+\,, \quad \widetilde{k}=k r_+\,, \quad \widetilde{\omega}=\omega r_+\,.
\end{equation}

\begin{figure}[th]
\centering
\includegraphics[width=.46\textwidth]{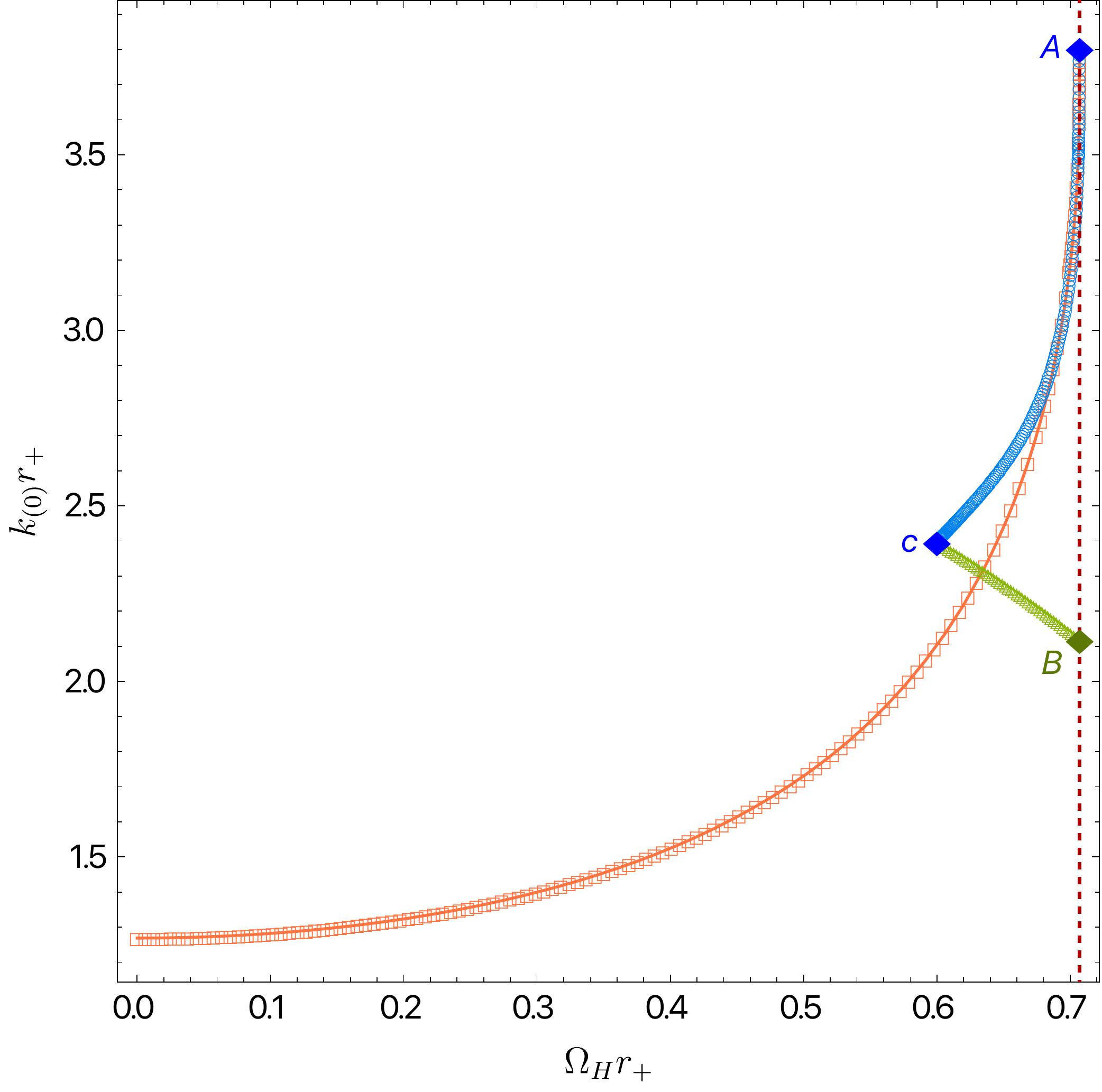}\hspace{1cm}
\includegraphics[width=.46\textwidth]{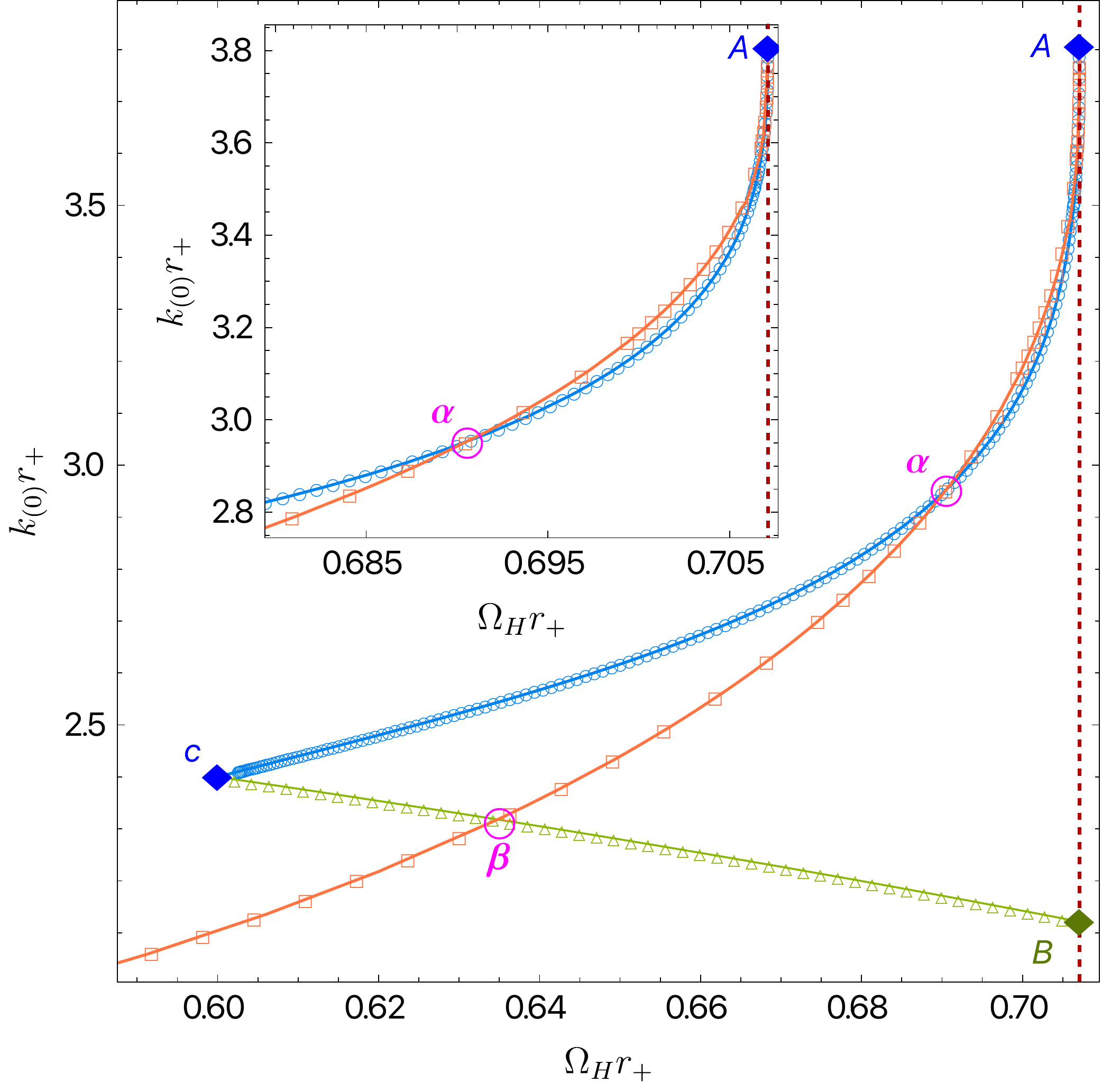}
\caption{Superradiant instability (for azimuthal number $m=2$) and Gregory-Laflamme instability of Myers-Perry black strings with parameters $\widetilde{k}_{(0)}$, $\widetilde{\Omega}_H$.  Superradiant instability occurs inside the triangular region $ABc$, and Gregory-Laflamme instability occurs below the curve marked by orange squares. The Gregory-Laflamme onset curve intersects with the edge of the unstable superradiant region at points $\alpha$ and $\beta$. The vertical dashed line at $\widetilde{\Omega}_H = 1/\sqrt{2}$ is extremality. 
} \label{Fig:zeroModeSuperGL}
\end{figure}

Fig.~\ref{Fig:zeroModeSuperGL} summarizes key information about the $m=2$ superradiant instability previously found in \cite{Dias:2022mde}.  In parameters $(\widetilde\Omega_H,\widetilde k)$, Myers-Perry black strings are unstable in a triangular region $ABc$ of Fig.~\ref{Fig:zeroModeSuperGL}.  This triangular region is defined by the region between two curves ($Bc$ and $Ac$) $\: \widetilde{k}_{\star}(\widetilde{\Omega}_H) \leq \widetilde{k}\leq \widetilde{k}_{(0)}(\widetilde{\Omega}_H)$ and extremality  ($AB$) $\widetilde{\Omega}_H \leq  1/\sqrt{2}\:$.  We also have the intersection point ($c$) of these curves $\widetilde{\Omega}_H|_c$, where $\: \widetilde{k}_{\star}(\widetilde{\Omega}_H|_c) =\widetilde{k}_{(0)}(\widetilde{\Omega}_H|_c)$.

Let us explain the physical origin of the curves $\widetilde{k}_{(0)}(\widetilde{\Omega}_H)$ and $\: \widetilde{k}_{\star}(\widetilde{\Omega}_H)$.  As we have alluded to earlier, the superradiant instability requires two ingredients.  One is the superradiant amplification of incident waves. The other is a confinement mechanism that causes the amplification process to continue growing until an instability sets in.  The curve $\widetilde{k}_{(0)}(\widetilde{\Omega}_H)$ is the superradiant onset curve where $\widetilde{\omega}=2m\widetilde{\Omega}_H$ with $\mathrm{Im}\,\widetilde{\omega}=0$. The curve $\: \widetilde{k}_{\star}(\widetilde{\Omega}_H)$ is the confining cutoff curve, where the confinement mechanism is no longer strong enough to support an instability.

It turns out that some parts of the unstable region $ABc$ of Fig.~\ref{Fig:zeroModeSuperGL} are known analytically for any $m$ \cite{Dias:2022mde}.  Point $A$, where the superradiant onset curve meets extremality $\widetilde{k}_{(0)}(\widetilde{\Omega}^{\mathrm{ext}}_H)$, is given by
\begin{equation}\label{OnsetR=Ext}
k_{(0)}^{(m)} r_+ |_{\hbox{\tiny ext}}=\sqrt{2 m^2 + 4m -1}\,.
\end{equation}
The confining cutoff curve ($Bc$ in Fig.~\ref{Fig:zeroModeSuperGL}) is given by
\begin{equation}\label{Cutoff}
\widetilde{k}_\star^{(m)} (\widetilde{\Omega}_H)=(2m-1)\sqrt{1-\widetilde{\Omega}_H^2}\,,
\end{equation}
and the intersection point between the superradiant onset curve and confining cutoff curve (point $c$ in Fig.~\ref{Fig:zeroModeSuperGL}) is
\begin{equation}\label{OmegaC}
\widetilde{\Omega}_H \big|_c=\frac{2 m-1}{\sqrt{8 m^2-4 m+1}}\,.
\end{equation}

For reference, the curve marked by orange squares in Fig.~\ref{Fig:zeroModeSuperGL} shows the zero mode curve $\widetilde{k}_{(0)}\big|_{\hbox{\tiny GL}}$ where the Gregory-Lafamme instability sets in.  This instability exists for any rotation, $0 \leq \widetilde{\Omega}_H \leq  1/\sqrt{2}$  and for $\widetilde{L}>\widetilde{L}_{(0)}\big|_{\hbox{\tiny GL}}=2\pi/\widetilde{k}_{(0)}\big|_{\hbox{\tiny GL}}$.  This curve intersects the onset and cutoff curves of the superradiant instability at points $\alpha$ and $\beta$ in Fig.~\ref{Fig:zeroModeSuperGL}.  Together, the figure shows that in different regions of parameter space either, none, or both instabilities can be present.

At the onset of an instability, perturbations neither grow nor decay exponentially. These perturbations instead move the solution towards other steady-state solutions.  Our focus in this paper is on the steady-state solutions that should branch from the onset of the $m=2$ superradiant instability, in a phase diagram of asymptotically  ${\cal M}^{1,4}\times S^1$ backgrounds with equal angular momenta\footnote{We focus on the superradiant onset curve $\widetilde{k}_{(0)}(\widetilde{\Omega}_H)$ and not the confining cutoff curve $\: \widetilde{k}_{\star}(\widetilde{\Omega}_H)$.  The reason is that we have well-defined, regular perturbative solutions to the former but not the latter.}.  From the nature of the perturbations, one can see that these solutions are not time independent (they are time-periodic), and neither axisymmetric nor translationally invariant along the string direction.  In \cite{Dias:2022mde}, we called these solutions {\it black resonator strings} by analogy to similar solutions that branch from superradiant instabilities in black holes \cite{Dias:2011at,Dias:2011tj,Dias:2015rxy,Ishii:2018oms,Ishii:2020muv,Ishii:2021xmn}. We will construct these solutions explicitly in the next few sections.  We will also determine whether the black resonator strings have higher or lower entropy (horizon area) than the corresponding Myers-Perry black string with the same length, energy, and angular momenta.  By the horizon area law, these entropy considerations will provide important information about the time evolution and ultimate endpoint of the superradiant instability.

Like the superradiant instability, the onset of the Gregory-Laflamme instability also leads to a new family of solutions: stationary but non-uniform black strings \cite{Kleihaus:2007dg}.  We leave the study of these non-uniform Myers-Perry black strings to future work, but lessons can be drawn from previous results on other black strings \cite{Gubser:2001ac,Harmark:2002tr,Kol:2002xz,Wiseman:2002zc,Kol:2003ja,Harmark:2003dg,Harmark:2003yz,Kudoh:2003ki,Sorkin:2004qq,Gorbonos:2004uc,Kudoh:2004hs,Dias:2007hg,Harmark:2007md,Wiseman:2011by,Figueras:2012xj,Dias:2017coo}.  From these results, we would expect the non-uniform black strings to have less entropy than the uniform solutions for a given energy and angular momenta.  We also expect that this non-uniform family connects to localized or caged black holes that have spherical (and not string-like) horizon topology.  A localised black hole is expected to be the entropically dominant solution, again for fixed energy and angular momenta.

\section{Ansatz and thermodynamics for black resonator strings \label{sec:ansatzThermo}}
\subsection{Symmetries of Myers-Perry black strings and its perturbations}\label{sec:sym}

The Myers-Perry black string \eqref{MPstring} is time-translation invariant, axisymmetric, and translationally invariant ($\partial_t$,  $\partial_\psi$, and $\partial_z$ are Killing fields).  A less obvious symmetry is an $SU(2)$, which we will now explain.

Recall that the isometry group of $S^3$ is $SO(4) \simeq SU(2)_L \times SU(2)_R$.  It follows that, in the zero rotation limit,  Myers-Perry strings (\ie Schwarzschild strings)  also have these isometries. The Killing vectors generating $SU(2)_L$  (denoted by $\xi_\imath$) and $SU(2)_R$ (denoted by $\bar{\xi}_\imath$) are, respectively, given by
\begin{subequations}
\begin{equation}
\label{lKVrKV}
\begin{cases}
 \xi_1 = \cos\phi\,\partial_\theta +
\frac{1}{2}\frac{\sin\phi}{\sin\theta}\,\partial_\psi -
\cot\theta\sin\phi\,\partial_\phi , & \\
\xi_2 = -\sin\phi\,\partial_\theta +
\frac{1}{2}\frac{\cos\phi}{\sin\theta}\,\partial_\psi -
\cot\theta\cos\phi\,\partial_\phi , & \\
\xi_3 = \partial_\phi , &
  \end{cases}
  \end{equation}
  and
\begin{equation}
\begin{cases}
\bar{\xi}_1 = -\sin(2\psi)\, \partial_\theta + \frac{\cos (2\psi)}{\sin \theta}\, \partial_\phi - \frac{1}{2}\cot \theta \cos (2\psi) \partial_\psi , & \\
\bar{\xi}_2 = \cos (2\psi) \,\partial_\theta + \frac{\sin (2\psi)}{\sin \theta} \,\partial_\phi - \frac{1}{2} \cot \theta \sin (2\psi)\, \partial_\psi , & \\
\bar{\xi}_3 = \frac{1}{2}\partial_\psi \,. &
  \end{cases}
\end{equation}
\end{subequations}
Note that $\bar{\xi}_\imath$ are the normalised dual vectors of $\sigma_\imath$:
$(\sigma_\imath)_\alpha (\bar{\xi}_\jmath)^\alpha=\delta_{\imath \jmath} \ (\alpha=\theta,\phi,\psi)$.
We can define the ``angular momentum'' operators
$L_\imath= i\xi_\imath$ and  $R_\imath= i \bar{\xi}_\imath $ which satisfy the commutation relations $[L_\imath,L_\jmath]=i \epsilon_{\imath\jmath k}L_k$ and $[R_\imath,R_\jmath]=-i \epsilon_{\imath\jmath k} R_k$ (where  $\epsilon_{\imath\jmath k}$ is the Cartesian Levi-Civita tensor).
Under $SU(2)_L$ and $SU(2)_R$, the 1-forms \eqref{Eulerforms} transform as
\begin{equation}
 \pounds_{L_\imath} \sigma_\jmath=0\ ,\qquad \pounds_{R_\imath} \sigma_\jmath = -i\epsilon_{\imath\jmath k}\sigma_k\ ,
\label{LRsigma}
\end{equation}
The invariance of the $S^3$ metric $\mathrm{d}\Omega_{S^3}^2=\frac{1}{4}(\sigma_1^2+\sigma_2^2+\sigma_3^2)$ under $SU(2)_L\times SU(2)_R$ can now be checked using \eqref{LRsigma}. It further follows from the first equation in \eqref{LRsigma} that $\sigma_\imath$ are invariant under $SU(2)_L$ (hence, they are often called $SU(2)$-invariant 1-forms).

From the second equation in \eqref{LRsigma} one concludes that $R_\imath$ generate the three-dimensional rotations of the ``vector'' $(\sigma_1,\sigma_2,\sigma_3)$. In particular, $R_3$ generates $U(1)_\psi \subset SU(2)_R$, which describes rotations in the $\sigma_1\sigma_2$-plane.  Thus, while Schwarzschild black strings preserve $SO(4) \simeq SU(2)_L \times SU(2)_R$, once we turn-on  rotation, Myers-Perry black strings \eqref{MPstring} preserve the $SU(2)_L$ subgroup but break $SU(2)_R$. Yet, $SU(2)_R$ is not completely broken since the
$U(1)_\psi \subset SU(2)_R$ symmetry generated by $R_3$ is preserved (indeed note that ${\mathrm d}s^2_{\mathbb{C}\mathrm{P}^1}$ is independent of $\psi$). That is to say, the isometry group of the equal-spinning Myers-Perry black string \eqref{MPstring} is $\mathbb{R}_t \times U(1)_z \times U(1)_\psi  \times SU(2)_L$.

How do perturbations at the onset of the superradiant instability (described in section~\ref{sec:MPinstabilities}) break these symmetries?  By definition, an onset has $\mathrm{Im}\,\widetilde \omega=0$, but these perturbations  have $\mathrm{Re}\,\widetilde \omega\neq 0$.  These perturbations are therefore time-periodic.  As these perturbations have $\widetilde k\neq0$, they also break translation invariance along $z$.  All $m\neq0$ perturbations also break axisymmetry about $\psi$.  However, $m=2$ are special in that they are the only nontrival superradiant perturbations that preserve  $SU(2)_L$.  This extra symmetry is why we focus only on the $m=2$ perturbations. 

Beyond linear order, these $m=2$ perturbations therefore extend towards a branch of solutions that break $\mathbb{R}_t$, $U(1)_z$ and $U(1)_\psi$, but are time periodic (preserving a helical Killing field $K=\partial_t+\Omega_H \partial_\psi  \equiv \partial_\tau$ that forms $\mathbb{R}_\tau$), and preserve $SU(2)_L$\footnote{It turns out that though $U(1)_z$ and $U(1)_\psi$ are broken, it is possible to find a special perturbation that preserves a $U(1)$ involving a linear combination of $\partial_z$ and $\partial_\psi$, \ie a helical symmetry.  The resulting non-uniform solutions were called {\it helical black strings} in \cite{Dias:2022mde}.  These solutions, though more symmetric, necessarily have nonzero horizon velocity in the $z$ direction (akin to ``boosted" black strings), so we do not study them here, but leave them for a forthcoming paper \cite{Dias:2023nbj}.}.  These are the solutions which we have called black resonator strings, and we aim to construct these solutions and study their properties in the rest of the manuscript.

\subsection{Ansatz for black resonator strings}\label{sec:ansatz}

The most general ansatz that can describe an asymptotically ${\cal M}^{1,4}\times S^1$ rotating string with the above $SU(2)_L$ symmetries, helical Killing vector field $\partial_\tau$, and zero horizon velocity in the string direction, can be written as
\begin{align}\label{ansatz}
{\mathrm ds}^2&=r_+^2\bigg\{ -\frac{y^2 \tilde{F} \,q_1}{\tilde{H}}\,{\mathrm d}\tau^2
+\frac{4 q_2}{(1-y^2)^4 \tilde{F}}\,{\mathrm d}y^2 +q_7\left(\frac{\widetilde{L}}{2}{\mathrm d}x+q_3 {\mathrm d}y\right)^2
\nonumber\\
&\qquad\quad\: +\frac{1}{(1-y^2)^2}
\bigg[ q_4 \tilde{H} \left(\frac{\Sigma_3}{2}+ \frac{y^2 \tilde{\Omega}}{\tilde{H}}
\left[ 1+ (1-y^2)^3 q_6\right]{\mathrm d}\tau \right)^2
+q_5 \frac{1}{4}\left( q_8 \Sigma_1^2 +\frac{\Sigma_2^2}{q_8}\right) \bigg] \bigg\},
\end{align}
where $q_j=q_j(y,x)$, for $j=1,2,3,\cdots, 8$, are eight functions of a radial coordinate $y$ and of a direction $x$ (the string, with physical length $L\equiv\widetilde{L} r_+$, extends non-uniformly along this direction), and
\begin{align}\label{ansatz2}
\tilde{F}(y)&=\left(2-y^2\right) \left(1-\frac{\widetilde{a} ^2}{1-\widetilde{a} ^2}(1-y^2)^2\right)\,,\nonumber\\
\tilde{H}(y)&=1+\frac{\widetilde{a} ^2}{1-\widetilde{a} ^2}(1-y^2)^4\,,\nonumber\\
\tilde{\Omega}(y)&=\widetilde{a}  \left(2-y^2\right) \left[1+\left(1-y^2\right)^2\right].
\end{align}
$\Sigma_1, \Sigma_2, \Sigma_3$ are the $SU(2)$ left-invariant 1-forms on $S^3$ defined as in \eqref{Eulerforms}, but with $\psi$ replaced by $\Psi$, where the relation between $\psi$ and $\Psi$ is given below in \eqref{MPintoNewAnsatz}. Furthermore, it will be convenient to redefine both $q_1$ and $q_7$ so that we can read the energy and tension by taking a single derivative with respect to $y$. We thus set
\begin{subequations}
\begin{equation}
q_1(x,y)=1+(1-y^2)\,\hat{q}_1(x,y)\,,\quad q_7(x,y)=1+(1-y^2)\,\hat{q}_7(x,y)
\end{equation}
and
\begin{equation}
q_i(x,y)=\hat{q}_i(x,y)\,, \quad i\neq1,7\,.
\end{equation}
\end{subequations}

In \eqref{ansatz}, the coordinate $x$ describes the direction along the string.  To obtain this coordinate, we started with a periodic coordinate $z$ with (dimensionful) spatial period $L\equiv \widetilde{L} r_+$, \ie  $z\sim z+L$.  But, because we seek solutions with a $\mathbb{Z}_2$ symmetry around $z=0$, we restrict our domain to $z\in[0,L/2]$. Finally, we can then introduce the dimensionless coordinate $x=\frac{2}{L}\,z$ which ranges between $x\in[0,1]$
.

To further understand the motivation for the {\it ansatz}  \eqref{ansatz} note that the Myers-Perry black string \eqref{MPstring}-\eqref{MPfns}  can be rewritten as  \eqref{ansatz} with  $q_{1,2,4,5,7,8}=1$ and $q_{3,6}=0$  after doing the coordinate and field redefinitions
\begin{align}\label{MPintoNewAnsatz}
& t=r_+ \tau\,, \qquad r=\frac{r_+}{1-y^2}\,,\qquad \psi=\Psi+\Omega_H\,r_+ \tau \,, \qquad z=\frac{1}{2}\,\widetilde{L} r_+ \,x\,;  \nonumber \\
& F=\left( 1-\frac{r_+}{r}\right) \tilde{F}\,,\qquad H=\tilde{H}\,,  \qquad
\Omega=-\frac{1}{r_+}\left( 1-\frac{r_+}{r}\right)\tilde{\Omega}+\frac{a}{r_+^2}\,\tilde{H}\,.
\end{align} 
The new coordinates $(\tau,y,\theta,\phi,\Psi,x)$ are dimensionless. The shift in the azimuthal coordinate $\psi$ ensures that \eqref{ansatz} describes solutions where the angular velocity at the horizon vanishes at the expense of being in a rotating frame at infinity. It follows that, in these new coordinates, the Killing horizon generator $K=\partial_t+\Omega_H \partial_\psi$ of the Myers-Perry black string becomes
\begin{equation}\label{SKV}
K=\partial_\tau \,.
\end{equation}
Thus, {\it ansatz} \eqref{ansatz} describes black string solutions with horizon at $y=0$, generated by \eqref{SKV}, with temperature $T_H=\frac{1}{2 \pi  r_+} \frac{1-2 \widetilde{a} ^2}{\sqrt{1-\widetilde{a} ^2}}$ (measured with respect to time $t$ since 
 $g_{AB} (\partial_t)^A (\partial_t)^B \to -1$
at the asymptotic region), and that asymptote to ${\cal M}^{1,4}\times S^1$ at $y=1$.

Note that the ansatz \eqref{ansatz} is completely general given our symmetry requirements. Indeed, the system has a nonlinear symmetry that leaves the metric invariant when we shift the Euler angle $\Psi \to \Psi+\Psi_0$, and there is a choice of $\Psi_0$ that sets the cross term $\sigma_1\sigma_2$ to zero.\footnote{Concretely, setting $\Psi=\psi-\frac{1}{4}\arctan\left(\frac{Q_2}{Q_1}\right)$ where $Q_{1,2}=Q_{1,2}(r,z)$ one has $\sqrt{Q_1^2+Q_2^2}\left( \Sigma_1^2-\Sigma_2^2\right)=Q_1\left(\sigma_1^2-\sigma_2^2\right)+2Q_2\sigma_1\sigma_2$.} After fixing $\Psi_0$, deformations along $\mathbb{C}\mathrm{P}^{1}$ are described by two remaining degrees of freedom given by the functions $q_5$ and $q_8$. 
 Moreover, because our metric fields depend on $x$ and $y$, we allow for a cross term $\mathrm{d}x \mathrm{d}y$ proportional to $q_3$ to avoid fixing the gauge.

Further notice that linearizing   \eqref{ansatz} about the Myers-Perry black string, after using \eqref{MPintoNewAnsatz}, one gets a $m=2$ perturbation that is described by \eqref{SRpert} with $Q\propto \delta q_8$ and $\omega=2m\Omega_H=4\Omega_H$. That is, this linearization yields the $m=2$ superradiant onset mode of the Myers-Perry black string.  Thus,  \eqref{ansatz} is a good ansatz to study the nonlinear back-reaction of the $m=2$ superradiant onset linear mode discussed previously. 

Generically, solutions of \eqref{ansatz} are not time independent nor axisymmetric since the associated $\mathbb{R}_t$ and $U(1)_\psi$ symmetries are broken in \eqref{ansatz} when $q_8\neq 1$.\footnote{This becomes evident when we return to the frame that does not rotate at infinite and rewrite
 \begin{equation*}
q_8 \Sigma_1^2+\frac{1}{q_8}\Sigma_2^2
 =\frac{1}{2} \left(q_8-\frac{1}{q_8}\right)
 \big[ \left(\sigma_1^2-\sigma_2^2\right) \cos (4 \Omega_H t )+2 \sigma_1 \sigma_2 \sin (4 \Omega_H t )\big]+\frac{1}{2} \left(q_8+\frac{1}{q_8}\right) \left(\sigma_1^2+\sigma_2^2\right)
\end{equation*}
which explicitly depends on $t$ when $q_8\neq 1$. Moreover, this term also depends explicitly on $\psi$, when  $q_8\neq 1$, as can be seen when we expand $\left(\sigma_1^2-\sigma_2^2\right)$ in terms of the Euler angles as in \eqref{SRpert}. Further recall that $\left(\sigma_1^2+\sigma_2^2\right)$ is the line element of $\mathbb{C}\mathrm{P}^{1}$ which is isometric to an $S^2$.} They clearly also break translation invariance along the string direction $x$. Thus, \eqref{ansatz} only has the symmetries $\mathbb{R}_\tau \times SU(2)_L$ with the $\mathbb{R}_\tau$ isometry generated by the horizon generator \eqref{SKV}.
 In this sense, we say that resonator solutions of \eqref{ansatz}  are time-periodic because they still have the helical Killing vector field \eqref{SKV}. Like black resonators in AdS, the helical Killing field is timelike in certain regions of the asymptotic AdS boundary but spacelike in others \cite{Dias:2011at,Dias:2015rxy,Ishii:2018oms,Ishii:2020muv,Ishii:2021xmn}.

To find the resonator string solutions, and to fix the remaining gauge freedom in our ansatz, we use the Einstein-DeTurck formalism \cite{Headrick:2009pv,Figueras:2011va,Wiseman:2011by,Dias:2015nua,Figueras:2016nmo}. This formulation of the gravitational equations requires a choice of reference metric $\overline g$, which has the same causal structure and contains the symmetries of the desired solution (it can have other symmetries).  The reference metric we choose is the Myers-Perry black string metric given by \eqref{ansatz} with  $q_{1,2,4,5,7,8}=1$ and $q_{3,6}=0$, as discussed above. The DeTurck method modifies the Einstein equation $R_{AB}=0$ into
\begin{equation}\label{EdeT}
R_{AB}-\nabla_{(A}\xi_{B)}=0\;,\qquad \xi^A \equiv g^{C D}[\Gamma^A_{\:C D}-\overline{\Gamma}^A_{\:C D}]\;,
\end{equation}
where $\Gamma$ and $\overline{\Gamma}$ define the Levi-Civita connections for $g$ and $\bar g$, respectively. Unlike $R_{\mu\nu}=0$, this equation yields a well-posed elliptic boundary value problem. Indeed, it was proved in \cite{Figueras:2011va} and \cite{Figueras:2016nmo} that static and stationary (with $t-\psi$ symmetry) solutions to \eqref{EdeT} necessarily satisfy $\xi^\mu=0$, and hence are also solutions to $R_{\mu\nu}=0$. Note that the results of \cite{Figueras:2011va,Figueras:2016nmo} apply to asymptotically flat and asymptotically AdS spacetimes and, of relevance here, to asymptotically Kaluza-Klein backgrounds.

We now discuss the boundary conditions. At the asymptotic boundary, $y=1$, we impose as a Dirichlet condition that our solutions must approach the reference metric. At $y=0$, we demand a regular bifurcate Killing horizon generated by $\partial_\tau$. This amounts to impose Neumann boundary conditions for $q_{1,2,4,5,6,7,8}$ and Dirichlet boundary condition for $q_3$. The discrete $\mathbb{Z}_2$ symmetry $x\to-x$ requires that all of the $q_i$'s have Neumann boundary conditions at $x=0$, except $q_3$ which has a Dirichlet boundary condition.  Finally, regularity at $x=1$ requires a Dirichlet boundary condition for $q_3$ and Neumann conditions for the remaining functions. 

We are now ready to solve the Einstein-DeTurck differential equations subject to the above boundary conditions. We will do this within higher-order perturbation theory about the Myers-Perry black string in section~\ref{sec:perturbative}. We will also solve the full nonlinear problem numerically. For that, we will use a Newton-Raphson algorithm and discretise the Einstein-DeTurck equations using pseudospectral collocation (with Chebyshev-Gauss-Lobatto nodes along the $x$ and $y$ directions). The resulting algebraic linear systems are solved by LU decomposition. These methods are reviewed in \cite{Dias:2015nua}.

It is convenient for us to work in units of the horizon radius $r_+$, where the Myers-Perry black string can be parametrized by the dimensionless quantities $\widetilde{a}=a/r_+$ and $\widetilde{L}=L/r_+$.  Later, we can convert these to units of $L$, as we will discuss in detail in section \ref{sec:thermo}.

The above discussions naturally invite us to follow one of two strategies (we will use both) to generate the 2-parameter space of black resonator strings:
\begin{enumerate}
\item We can choose to generate lines of black resonator strings that have the same dimensionless rotation $\widetilde{a}$ as the Myers-Perry black string they branch from. The dimensionless length $\widetilde{L}$ is varied along these families of solutions.
\item Alternatively, we can choose to generate lines of black resonator strings that have the same dimensionless length $\widetilde{L}$ as the Myers-Perry black string they branch from (\ie  we fix the length to be $\widetilde{L}=2\pi /\widetilde{k}_{(0)}$ where $\widetilde{k}_{(0)}$ is the $m=2$ zero mode wavenumber for the superradiant instability of the Myers-Perry black string that given in Fig.~\ref{Fig:zeroModeSuperGL}).
\end{enumerate}
Altogether, with these two strategies we span the 2-dimensional phase space parameter of black resonator strings. We will present our results in section~\ref{sec:PhaseDiag}.

\subsection{Asymptotic expansion}\label{sec:expansion}

Before computing the asymptotic charges, we need to understand the expansion of the functions $q_i$ near asymptotic infinity. This expansion can be sorted out via a generalised Frobenius expansion near asymptotic infinity, since the equations of motion linearise there. We find
\begin{subequations}
\begin{equation}
\hat{q}_i(x,y) = \sum_{n=0}^{+\infty}\hat{q}^{(n)}_i(y)\;\cos(n\,\pi\,x)\,,\quad i = 1,2,4,5,6,7\,,
\end{equation}
\begin{equation}
\hat{q}_3(x,y) = \sum_{n=0}^{+\infty}\hat{q}^{(n)}_3(y)\;\sin(n\,\pi\,x)\,,
\end{equation}
\begin{equation}
\hat{q}_8(x,y) =1+\sum_{n=1}^{+\infty}\hat{q}^{(n)}_8(y)\;\cos(n\,\pi\,x)\,.
\end{equation}
\end{subequations}

The exact decay of each of the $\hat{q}^{(n)}_i(y)$ is intricate, but from the above considerations it is clear that the only components that have a chance to contribute to the asymptotic charges are the $n=0$ components of the $i = 1,2,4,5,6,7$ functions. These turn out to be relatively easy to determine:
\begin{subequations}
\begin{equation}
\hat{q}^{(0)}_i(y)=\sum_{j=0}^{+\infty}\alpha_i^{(j)}(1-y)^j+(1-y)^4\log(1-y)\sum_{j=0}^{+\infty}\beta_i^{(j)}(1-y)^j
\end{equation}
with
\begin{equation}
\alpha_1^{(0)}=\alpha_2^{(0)}=\alpha_4^{(0)}=\alpha_5^{(0)}=\alpha_6^{(0)}=1
\end{equation}
and
\begin{equation}
\alpha_3^{(0)}=\alpha_6^{(0)}=0\,.
\end{equation}
\end{subequations}
\subsection{Thermodynamics for black resonator strings}\label{sec:thermo}

In this section, we find the thermodynamic quantities that will allow us to discuss the phase diagram of black resonator strings and compare them against the Myers-Perry black strings.

Consider the Einstein-Hilbert action with the Gibbons-Hawking term for an asymptotically $\mathcal{M}^{1,4}\times S^1$ spacetime manifold $\mathcal{M}$ with spacelike boundary $\partial \mathcal{M}$
\begin{equation}\label{RegAction}
S_{\rm reg}=-\frac{1}{16\pi G_6}\int_{\mathcal{M}} \mathrm{d}^6 x \sqrt{-g} \,R -\frac{1}{8\pi G_6}\int_{\mathcal{\partial M}} \mathrm{d}^5 x \sqrt{-h}\, K \,,
\end{equation}
where $G_6$ is Newton's constant, $R$ is the 6-dimensional Ricci scalar of the metric $g_{AB}$, $\sqrt{-g}$ is the determinant of $g_{AB}$,   $\sqrt{-h}$ is the determinant of the pullback of $h^{AB}=g^{AB}-n^{A} n^{B}$ into the 5-dimensional boundary $\partial \mathcal{M}$, $n$ is the unit normal to $\partial \mathcal{M}$ with $n_\mu n^{\mu}=1$,  and  $K=g^{AB}K_{AB}$ is the trace of the extrinsic curvature $K_{AB}=h_A^{\phantom{A} C}\nabla_C n_B$ of $\partial \mathcal{M}$. The addition of the Gibbons-Hawking term guarantees that one gets a regularized action (\ie  with a well-defined variational principle) since the variation \eqref{RegAction} w.r.t. $g^{AB}$ yields the correct Einstein bulk equation of motion, $R_{AB}-\frac{1}{2}R g_{AB}=0$. But, using this Einstein bulk equation, the variation of \eqref{RegAction} still yields the boundary term (with $a,b$ running over the boundary coordinates)
\begin{equation}\label{VarRegAction}
\delta S_{\rm reg}= -\frac{1}{16\pi G_6} \int_{\partial \mathcal{M}} \mathrm{d}^5 x \sqrt{-h} \left( K_{ab} -K h_{ab} \right) \delta h^{ab}\,,
\end{equation}
from which the regularized energy-momentum tensor becomes
\begin{equation}\label{RegTensor}
T_{ab}^{\rm reg}=-\frac{2}{\sqrt{-h}} \frac{\delta S_{\rm reg} }{\delta h^{ab}}=\frac{1}{8\pi G_6}\left( K_{ab} -K h_{ab} \right).
\end{equation}
Unfortunately, this stress tensor diverges as we approach the asymptotic boundary.  To get a renormalized action we can  add counterterm contributions to $S_{\rm reg}$ that only depend on the boundary metric, (and thus preserve the bulk equations of motion) that annihilate these divergences. Following the renormalization procedure of \cite{Kraus:1999di} (see also \cite{Mann:2005yr,Kleihaus:2009ff}) one finds that renormalization is achieved with the following counterterm\footnote{For asymptotically $\mathcal{M}^{1,D-p-1}\times \mathbb{T}^p$ spacetimes with $p$ brane directions, we find that the appropriate counterterm is $S_{\rm ct}= -\frac{1}{8\pi G_D} \int_{\partial \mathcal{M}} \mathrm{d}^{D-1} x \sqrt{-h} \sqrt{C_{(D,p)}\mathcal{R}}$ with $C_{(D,p)}=\sqrt{\frac{D-p-2}{D-p-3}}$. In our case $D=6,p=1$.}
\begin{equation}\label{CountertermAction}
S_{\rm ct}= -\frac{1}{8\pi G_6} \int_{\partial \mathcal{M}} \mathrm{d}^5 x \sqrt{-h} \sqrt{\frac{3}{2}\mathcal{R}}\,,
\end{equation}
where $\mathcal{R}$ is the Ricci scalar of the boundary metric $h_{ab}$. Its variation yields
\begin{equation}\label{varCountertermAction}
\delta S_{\rm ct}= -\frac{1}{8\pi G_6} \int_{\partial \mathcal{M}} \mathrm{d}^5 x \sqrt{-h}\, \frac{1}{2}\sqrt{\frac{3}{2\,\mathcal{R}}}\left( \mathcal{R}_{ab}-\mathcal{R} \,h_{ab}\right)\delta h^{ab}\,,
\end{equation}
where $ \mathcal{R}_{ab}$ is the Ricci tensor of the boundary metric. Therefore, the renormalized energy-momentum stress tensor that follows from variation of the renormalized action $S_{\rm ren}=S_{\rm reg}+S_{\rm ct}$ is
\begin{equation}\label{RenTensor}
T_{ab}^{\rm ren}=\frac{1}{8\pi G_6}\left( K_{ab} -K h_{ab} + \sqrt{\frac{3}{2\,\mathcal{R}}} \,\mathcal{R}_{ab}-\sqrt{\frac{3}{2}\mathcal{R}} \, h_{ab} \right).
\end{equation}
The conserved charge associated to an asymptotic Killing vector field $\xi$ at the asymptotic boundary $\Sigma_t$ (a constant time slice $t=$constant at $y=1$)  is then
\begin{equation}\label{charge}
Q_\xi=\lim_{y\to 1}\int_{\Sigma_t} d^4x\sqrt{\sigma} \,\eta^i \,\xi^j \,T_{ij}^{\rm ren}.
\end{equation}
where $\eta$ is the unit normal $\eta_i=\frac{\partial_i \mathfrak{f}}{|\partial \mathfrak{f} |}$ (with $\eta_i \eta^i=-1$) to the spacelike hypersurface $\Sigma_t$ defined by $\mathfrak{f}=0$, and $\sqrt{\sigma}$ is the determinant of $\sigma^{ij}$ which is the pullback of $\sigma^{ab}=h^{ab}+\eta^{ab}$ to $\Sigma_t$.
It follows that the energy of the system associated to the asymptotic Killing vector field $\xi=\partial_t=\partial_\tau -\Omega_H \partial_\Psi$ is given by $E\equiv Q_{\partial_t}$ and that the angular momentum is the conserved charge $J\equiv Q_{\partial_\psi}$ associated to the asymptotic Killing vector field $\xi=\partial_\psi=\partial_\Psi$. Finally, the tension associated to the asymptotic Killing vector field $\xi=\partial_z$ is $T_z\equiv Q_{\partial_z}$. Altogether, the energy, angular momentum and tension of the black resonator strings are:
\begin{align}\label{chargesResonator}
&E=\frac{\pi  \,\widetilde{L} \,r_+^3}{16 G_6} \left(\frac{6}{1-\widetilde{a} ^2}-3\alpha_1^{(1)}-\alpha_7^{(1)}\right), \nonumber \\
& J=\frac{\pi  \,\widetilde{L} \,r_+^4}{8 G_6}\,\widetilde{a} \left(\frac{2}{1-\widetilde{a} ^2}-\alpha_6^{(1)}\right), \nonumber\\
& T_z=\frac{\pi \, r_+^2}{16 G_6}\left(\frac{2}{1-\widetilde{a} ^2}-\alpha_1^{(1)}-3\alpha_7^{(1)}\right).
\end{align}
The temperature (w.r.t. the asymptotic Killing vector field $\partial_t$ with normalization 
$\lVert|\partial_t\rVert|^2\to-1$
at the asymptotic region), the angular velocity and the entropy of the black resonator strings are
\begin{equation}
\label{TOSResonator}
T_H=\frac{1}{2 \pi r_+} \frac{1-2 \widetilde{a} ^2}{\sqrt{1-\widetilde{a} ^2}}\,,\quad
 \Omega_H=\frac{\widetilde{a} }{r_+}\,,\quad
 S_H=\frac{\pi ^2 \,\widetilde{L} \, r_+^4}{2 G_6\,\sqrt{1-\widetilde{a} ^2}}\int_0^1 \sqrt{q_4(x,0)} q_5(x,0) \sqrt{q_7(x,0)}\,\mathrm{d}x\,.
\end{equation}
To express quantities in units of the Kaluza-Klein circle length $L$, we will use the following dimensionless energy, angular momentum, tension, temperature,  angular velocity and entropy:
\begin{equation}\label{dimensionlessThermo}
\mathcal{E}\equiv \frac{E}{L^3}\,, \qquad \mathcal{J}\equiv \frac{J}{L^4}\,, \qquad \mathcal{T}_z\equiv \frac{T_z}{L^2}\,, \qquad  \tau_H \equiv T_H L\,, \qquad  \omega_H\equiv \Omega_H L \,, \qquad \sigma_H\equiv \frac{S_H}{L^4}
\end{equation}
to discuss our phase diagram of asymptotically $\mathcal{M}^{1,4}\times S^1$ solutions.

Next, we would like to find the Smarr and first law for these scale invariant quantities.\footnote{The first law for the Schwarzschild black string was discussed in \cite{Townsend:2001rg,Harmark:2003eg,Kastor:2007wr}.} For this purposes, first notice that the extensive thermodynamic variables of the system are the total energy, $E$, angular momentum $J$, tension $T_z$, the total entropy $S_H$, and the length $L$. It follows that the first law for the total charges of the system is:
\begin{equation}\label{1st}
\dd E= T_H \,\dd S_H+  2\,\Omega_H \dd J + T_z\, \dd L  \,,
\end{equation}
where the factor of $2$ accounts for the fact that the two angular momenta of our solutions are equal, and we see that $T_H, \Omega_H, T_z$ are the  potentials (intensive variables) conjugate to $S_H, J, L$, respectively. Under the scaling symmetry
\begin{align}\label{scalingSym}
& \{\tau,y,\theta,\phi,\Psi\}\to \{\tau,y,\theta,\phi,\Psi\},\qquad \{r_+,a,L\}\to \left\{\lambda r_+, \lambda\, a, \lambda L \right\}, \qquad \nonumber\\
&\{ q_{j} \}\to \{ q_{j} \}, (j=1,\cdots,8)\,
\end{align}
the energy scales as $E\to \lambda^3 E$ and thus it is a homogeneous function of $\lambda$ of degree 3. This means that for any value of $\lambda$,
\begin{equation}\label{Euler1}
E\left(\lambda^4 \,S_H,\lambda^4 \,J, \lambda\,L\right)= \lambda^3 \,E\left(S_H, J, L\right).
\end{equation}
We can now apply Euler's theorem for homogeneous functions to write the energy as a function of its partial derivatives:\footnote{Essentially, in the present case, Euler's theorem amounts to taking a derivative of the homogeneous relation \eqref{Euler1} with respect to $\lambda$ and then sending $\lambda\to 1$.}
\begin{equation}\label{Euler2}
4 S_H \,\frac{\partial E}{\partial S_H} + 4 J\, \frac{\partial E}{\partial J}+ L \frac{\partial E}{\partial L}
=3 E\left(S_H, \,J, \,L\right).
\end{equation}
Reading the partial derivatives in \eqref{Euler2} from \eqref{1st}, one gets the Smarr relation for the charges of the system:
\begin{equation}\label{Smarr}
 E=\frac{4}{3}\left( T_H \,S_H +  2\,\Omega_H \, J + \frac{1}{4}T_z \,L \right)\,.
\end{equation}
It will be useful to have also the Smarr relation and first law for the dimensionless quantities \eqref{dimensionlessThermo}. To get the former, we simply need to divide  \eqref{Smarr} by $L^3$ yielding
\begin{equation}\label{SmarrDless}
 \mathcal{E}=\frac{4}{3}\left( \tau_H \,\sigma_H +  2\,\omega_H \, \mathcal{J} + \frac{1}{4}\mathcal{T}_z  \right)\,.
\end{equation}
Finally, we can rewrite the first law \eqref{1st} in terms of \eqref{dimensionlessThermo}  and use  \eqref{Smarr}  to find the desired first law for the dimensionless quantities
\begin{equation}\label{1stLawDless}
\dd  \mathcal{E}=  \tau_H\,\dd \sigma_H+ 2\,\omega_H \, \dd \mathcal{J}\,,
\end{equation}
which, as expected, does not include a contribution proportional to the quantity $L$ we use for our units. In a traditional thermodynamic language, the first law \eqref{1stLawDless} and the Smarr relation \eqref{SmarrDless} are also known as the Gibbs-Duhem and Euler relations, respectively.

From \eqref{chargesResonator}-\eqref{dimensionlessThermo} the dimensionless thermodynamic quantities read
\begin{align}\label{ThermoResonatorDless}
&\mathcal{E}=\frac{\pi}{16 G_6 \widetilde{L}^2} \left(\frac{6}{1-\widetilde{a} ^2}-3\alpha_1^{(1)}-\alpha_7^{(1)}\right), \nonumber \\
& \mathcal{J}=\frac{\pi}{8 G_6\widetilde{L}^3}\,\widetilde{a} \left(\frac{2}{1-\widetilde{a} ^2}-\alpha_6^{(1)}\right), \nonumber\\
& \mathcal{T}_z=\frac{\pi}{16 G_6 \widetilde{L}^2}\left(\frac{2}{1-\widetilde{a} ^2}-\alpha_1^{(1)}-3\alpha_7^{(1)}\right)\nonumber\\
&\tau_H=\frac{\widetilde{L}}{2 \pi} \frac{1-2 \widetilde{a} ^2}{\sqrt{1-\widetilde{a} ^2}}\,,\nonumber \\
& \omega_H=\widetilde{a} \,\widetilde{L}\,,\nonumber \\
& \sigma_H=\frac{\pi ^2}{2 G_6 \widetilde{L}^3}\frac{1}{\sqrt{1-\widetilde{a} ^2}} \int_0^1\sqrt{q_4(x,0)} q_5(x,0) \sqrt{q_7(x,0)}\,\mathrm{d}x
\end{align}
which will be used to discuss our physical results.

We will use \eqref{SmarrDless}-\eqref{1stLawDless} to check our results. Note that when we set $\alpha_i^{(1)}=0$ in \eqref{ThermoResonatorDless}, we recover the dimensionless quantities associated to the thermodynamics \eqref{ThermoMP} of the Myers-Perry black string. 

\section{Perturbative construction of resonator strings \label{sec:perturbative}}

We are now in a position to construct black resonator strings and study their thermodynamics.  Here, we begin with a perturbative construction, which describes black resonator strings in the region near their merger with the Myers-Perry black strings.  We do this by solving the boundary vale problem described in the previous section to fifth order in perturbation theory.

We follow a perturbative approach developed in \cite{Dias:2017coo,Bea:2020ees}, which originated from \cite{Gubser:2001ac,Wiseman:2002zc,Sorkin:2004qq}.  Using linear perturbation theory, we first identify the region of parameter space where Myers-Perry black strings are unstable to $m=2$ superradiant modes. This task was already performed in \cite{Dias:2022mde} and explained in section \ref{sec:MPinstabilities} and shown in Fig.~\ref{Fig:zeroModeSuperGL}, but we will shortly describe this calculation in the present context. Our main task here is to continue the perturbative expansion to higher orders, until we reach an order where thermodynamic quantities receive perturbative corrections.

Let us now describe our perturbative analysis.  We work with the ansatz \eqref{ansatz} and equations of motion \eqref{EdeT} in the previous section, and write the following expansion
\begin{subequations} \label{PTexpansion}
\begin{eqnarray}
&& q_j(x,y)=\mathcal{Q}_j+\sum_{n=1}^{\infty} \epsilon^n \,q_j^{(n)}(x,y); \label{PTexpansionA} \\
&& \widetilde{k}=\sum_{n=1}^{\infty} \epsilon^{n-1} \widetilde{k}^{(n-1)}\equiv \widetilde{k}_{(0)}+\sum_{n=2}^{\infty} \epsilon^{n-1} \widetilde{k}^{(n-1)},\quad \hbox{with} \:\: \widetilde{L}=\frac{2\pi}{\widetilde{k}}, \label{PTexpansionB}
\end{eqnarray}
\end{subequations}
where $\mathcal{Q}_{1,2,4,5,7,8}=1$, $\mathcal{Q}_{3,6}=0$ describes the Myers-Perry black string background.  Note
that the equations of motion depend explicitly on $L = 2\pi/k$,  so both $k$ and $L$ receive perturbative corrections, the latter of which we write as
\begin{equation}
\widetilde{L}= \widetilde{L}_{(0)}+\sum_{n=2}^{\infty} \epsilon^{n-1}  \widetilde{L}^{(n-1)}\,,
\end{equation}
and its expansion coefficients can be read straightforwardly once \eqref{PTexpansionB} is known.\footnote{\label{footScheme}We have chosen to work with a coordinate $x\in [0,1]$ rather than $z=x \frac{L}{2}\in [0, \frac{L}{2}]$.  The Fourier modes between these coordinates are related (for integer $\eta$) as $\cos(\eta \,k \,z)=\cos(\eta\, \pi \,x)$.  We also have a factor of $\pi$ in the cosine instead of $2\pi$ in order to make use of the $\mathbb{Z}_2$ symmetry in the solutions we seek.
}

Our expansion parameter $\epsilon$ is the amplitude of the linear order perturbation ($n=1$). We want to consider a superradiant perturbation of the form \eqref{SRpert} for $m=2$ at its onset, i.e. with $\omega=2m\Omega_H=4\Omega_H$.  Such a perturbation breaks the $\mathbb{R}_t$, $U(1)_\psi$, and $U(1)_x$ symmetries of the Myers-Perry black string. For our ansatz, the only metric component that is perturbed at linear order is $q_8(x,y)$ with a deformation of the form
\begin{equation}\label{PT:n1}
q_8^{(1)}(x,y)=\mathfrak{q}_8^{(1)}(y)\cos(\pi\, x)\,,\qquad q^{(1)}_{1,2,3,4,5,6,7}(x,y)=0.
\end{equation}
The linearized equation of motion for $\mathfrak{q}_8^{(1)}(y)$  is a quadratic eigenvalue problem for $\widetilde{\kappa}\equiv  \sqrt{\widetilde{k}_{(0)}^2-\widetilde{\Omega}_H^2}$ that we solve to get the superradiant onset wavenumber $\widetilde{k}_{(0)}$. This task was already done in \cite{Dias:2022mde} and shown as the  blue curve $cA$ in Fig.~\ref{Fig:zeroModeSuperGL}. Linear analysis also identifies the region $ABc$ of parameter space where Myers-Perry black strings are unstable to $m=2$ superradiant modes (Fig.~\ref{Fig:zeroModeSuperGL}).

To fix the normalization condition for the linear problem (which also fixes the expansion parameter of our perturbation scheme), we define
\begin{equation}
\mathfrak{q}_8^{(1)}(y)=\left(1-y^2\right)^{3/2} e^{-\sqrt{\widetilde{k}_{(0)}^2-\widetilde{\Omega}_H^2}/(1-y^2)} \widehat{\mathfrak{q}}_8(y)
\end{equation}
and require that $ \widehat{\mathfrak{q}}_8|_{y=1}\equiv 1$.  This fixes the horizon value $\widehat{\mathfrak{q}}_8|_{y=0}\equiv \mathfrak{q}_8^{H}$ and thus the horizon function  $\mathfrak{q}_8^{(1)}|_{y=0} \cos(\pi x)\equiv \mathfrak{q}_8^{H}\cos(\pi x)$.  When we continue with perturbation theory at higher order, we will require that the $\cos(\pi x)$ component of $q_8$ at the horizon does not receive any perturbative corrections.  This uniquely fixes the expansion parameter $\epsilon$.

Although the sector of perturbations we look at only excites  $q_8(x,y)$ at linear order $-$ see \eqref{PT:n1} $-$  at higher order this backreacts on all other metric components, whose perturbations are  described by \eqref{PTexpansion}. More concretely, at higher orders $\mathcal{O}(\epsilon^n)$, the expansion \eqref{PTexpansion} gives a boundary-value problem for the quantities $\{ \widetilde{k}^{(n-1)}, q_j^{(n)}\}$. Note again that the wavenumber $\widetilde{k}$ and thus the length $\widetilde{L}$ are also corrected at higher orders.  Since we have fixed the $x$ coordinate to lie in $x\in [0,1]$, we can express the $x$-dependence of the functions as a sum of Fourier modes (with harmonic number $\eta$)
\begin{equation} \label{FourierExp}
q_j^{(n)}(x,y)=\sum_{\eta=0}^{n} \mathfrak{q}_j^{(n,\eta)}(y) \cos(\eta \,\pi \,x).
\end{equation}
Here and onwards, $\eta=0,\ldots, n$ identifies a particular Fourier mode of our expansion at order $\mathcal{O}(\epsilon^n)$ and we have the identification $\mathfrak{q}_j^{(1,1)}\equiv \mathfrak{q}_j^{(1)}$ with the latter function introduced in \eqref{PT:n1}. That is, we have already found the $n=1$ contribution, \eqref{PT:n1} and $\widetilde{k}^{(0)}\equiv \widetilde{k}_{(0)}$, of this expansion by solving a homogeneous eigenvalue problem. At linear order, we started with the single $\eta=1$ Fourier mode and the $n^{\rm th}$ polynomial power of this linear mode has the highest harmonic $\eta=n$.  This implies that the Fourier series at order $n$ terminates at harmonic $\eta=n$. 

Let us now describe the structure of the perturbative equations at order $\mathcal{O}(\epsilon^n)$, $n\geq 2$ in full detail.  Because the Fourier mode $\eta=1$ is the only one that exists at linear order, the structure of the $\eta=1$ equations are different from the rest.  

So here, we begin first with the more general $\eta\neq1$ case. Note that for $n\ge 2$, the perturbative equations of motion are no longer homogeneous, but is now a boundary value problem with a source.  The perturbative equations of motion take the form\footnote{Because only $ \mathfrak{q}_8^{(1)}$ is excited at linear order, the ODE for $\mathfrak{q}_8^{(n,\eta)}$ decouples from the others in \eqref{highEoMgeneral}. That is, \eqref{highEoMgeneral} is effectively a coupled system of ODEs for $\mathfrak{q}_{j\leq 7}^{(n,\eta)}$ plus a decoupled ODE for $\mathfrak{q}_8^{(n,\eta)}$.}
\begin{equation}\label{highEoMgeneral}
\bar{{\cal L}}_H \, \mathfrak{q}_j^{(n,\eta)}={\cal S}_j^{(n,\eta)}, \quad \hbox{if $n\geq 2$ and $\eta\neq 1$,}
\end{equation}
where the source ${\cal S}_j^{(n,\eta)}$ is a function of the lower order solutions and their derivatives, and where $\bar{{\cal L}}_H$ is a differential operator that only depends on the background Myers-Perry solution $\mathcal{Q}_j$ and superradiant onset wavenumber $\widetilde{k}_{(0)}$. 
The general solution $\mathfrak{q}_j^{(n,\eta)}(y)$ is found by solving \eqref{highEoMgeneral} subject to the boundary conditions we already discussed below \eqref{EdeT}.\footnote{That is, we require vanishing asymptotic Dirichlet boundary conditions $\mathfrak{q}_j^{(n,\eta)}|_{y=1}=0$ $-$ so that the full solution \eqref{PTexpansion} approaches the DeTurck reference Myers-Perry string solution, and regularity at the horizon $y=0$. The latter requires that we give a Dirichlet boundary condition for $ \mathfrak{q}_3^{(n,\eta)}$ and  Neumann boundary conditions for $\mathfrak{q}_{j\neq 3}^{(n,\eta)}$.
}

Moving on to the `exceptional' case $\eta=1$, at order $\mathcal{O}(\epsilon^n)$, $n\geq 2$, our boundary value problem becomes a (non-conventional) eigenvalue problem in $\widetilde{k}^{(n-1)}$. The 7 ODEs for $\mathfrak{q}_{j\leq 7}^{(n,1)}$ are still of the form \eqref{highEoMgeneral} and independent of $\mathfrak{q}_{8}^{(n,1)}$ and $\widetilde{k}^{(n-1)}$. The equation for $\mathfrak{q}_{8}^{(n,1)}$ decouples from the others (so it is independent of  $\mathfrak{q}_{j\leq 7}^{(n,1)}$) but it does depend on the eigenvalue  $\widetilde{k}^{(n-1)}$. More explicitly, the structure of the equations is \footnote{It is not a standard eigenvalue problem because the eigenvalue $\widetilde{k}^{(n-1)}$ is not multiplying the unknown eigenfunction $\mathfrak{q}_8^{(n,1)}$. Instead, it multiplies an eigenfunction $\mathfrak{q}_8^{(1)}$ that was already determined at previous $n=1$ order.}
\begin{eqnarray}\label{highEoMspecial}
&& {\cal L}_H \, \mathfrak{q}_{1,\cdots,7}^{(n,1)}={\cal S}_{1,\cdots,7}^{(n,1)}\,, \nonumber\\
&& {\cal L}_{H,8} \, \mathfrak{q}_8^{(n,1)}=\widetilde{k}^{(n-1)}\frac{8 \widetilde{k}_{(0)}  \mathfrak{q}_8^{(1)} }{(1-y^2)^4 f} + {\cal S}_8^{(n,1)}, \quad \hbox{if $n\geq 2$ and $\eta = 1$,}
\end{eqnarray}
where $f(y)$ was defined in \eqref{ansatz2},  $\mathfrak{q}_8^{(1)}(y)$ was introduced in \eqref{PT:n1}, and ${\cal L}_H$, ${\cal L}_{H,8}$ are the differential operators of the homogeneous equations. We now have to solve \eqref{highEoMspecial} (subject to boundary conditions that are the same as for the $\eta\neq 1$ case) to find $\mathfrak{q}_{j\leq 7}^{(n,1)}(y)$, the eigenvalue $\widetilde{k}^{(n-1)}$ and $\mathfrak{q}_8^{(n,1)}(y)$.

As it turns out (after doing the computation), not all Fourier modes $\eta=0,\ldots\,n$ are excited. For even $n$, only modes with even $\eta$ are excited.  Likewise, for odd $n\geq 3$, only modes with odd $\eta$ are excited. It follows that, to order $n=5$, the modes that are excited in our system are:
\begin{subequations}\label{excitedFourierModes}
\begin{align}
& q_j^{(2)}(x,y)=\mathfrak{q}_j^{(2,0)}(y)+\mathfrak{q}_j^{(2,2)}(y)\cos(2\,\pi \,x),  \\
& q_j^{(3)}(x,y)=\mathfrak{q}_j^{(3,1)}(y)\cos(\pi \,x)+\mathfrak{q}_j^{(3,3)}(y)\cos(3\,\pi \,x), \\
& q_j^{(4)}(x,y)=\mathfrak{q}_j^{(4,0)}(y)+\mathfrak{q}_j^{(4,2)}(y)\cos(2\,\pi \,x)+\mathfrak{q}_j^{(4,4)}(y)\cos(4\,\pi \,x),  \\
& q_j^{(5)}(x,y)=\mathfrak{q}_j^{(5,1)}(y)\cos(2\,\pi\, x)+\mathfrak{q}_j^{(5,3)}(y)\cos(3\,\pi \, x)+\mathfrak{q}_j^{(5,5)}(y)\cos(5\,\pi\, x).
\end{align}
\end{subequations}
This last observation, together with the previous observation that Fourier modes with $\eta=1$ are those that give the wavenumber correction $\widetilde{k}^{(n-1)}$ at order $\mathcal{O}(\epsilon^n)$, implies that $\widetilde{k}^{(n-1)}=0$ if $n$ is even. Furthermore, at even $n$ order, the $\cos{(\pi\, x)}$ Fourier mode is not excited by the source and thus the only solution of \eqref{highEoMspecial} is the trivial solution.

Finally, note that the $\eta=0$ harmonics are of particular special interest. Indeed, as we will see later in \eqref{EntropyDiff2}, modes with $\eta\neq 0$ do not directly contribute to the final thermodynamic quantities, though we still need to find the $\eta\neq 0$ harmonics at lower order to obtain the $\eta=0$ solution at order $n$ and the wavenumber corrections $\widetilde{k}^{(n-1)}$ (the latter also contribute to the above thermodynamic quantities). Moreover, it follows from the discussion of \eqref{excitedFourierModes} that odd order $n$ modes do not contribute to corrections of thermodynamic quantities.

\begin{figure}[th]
\centering
\includegraphics[width=.46\textwidth]{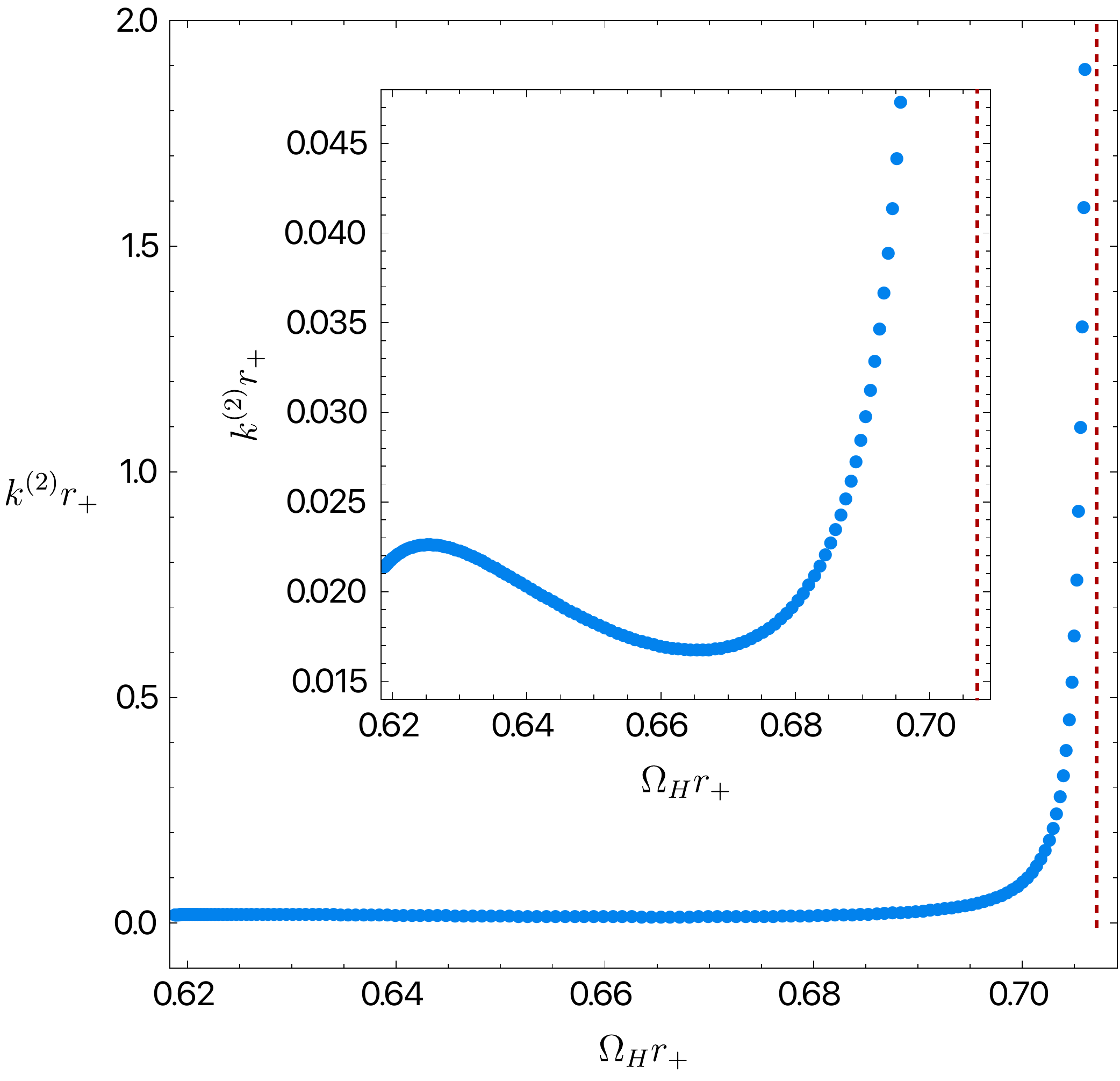}\hspace{0.8cm}
\includegraphics[width=.469\textwidth]{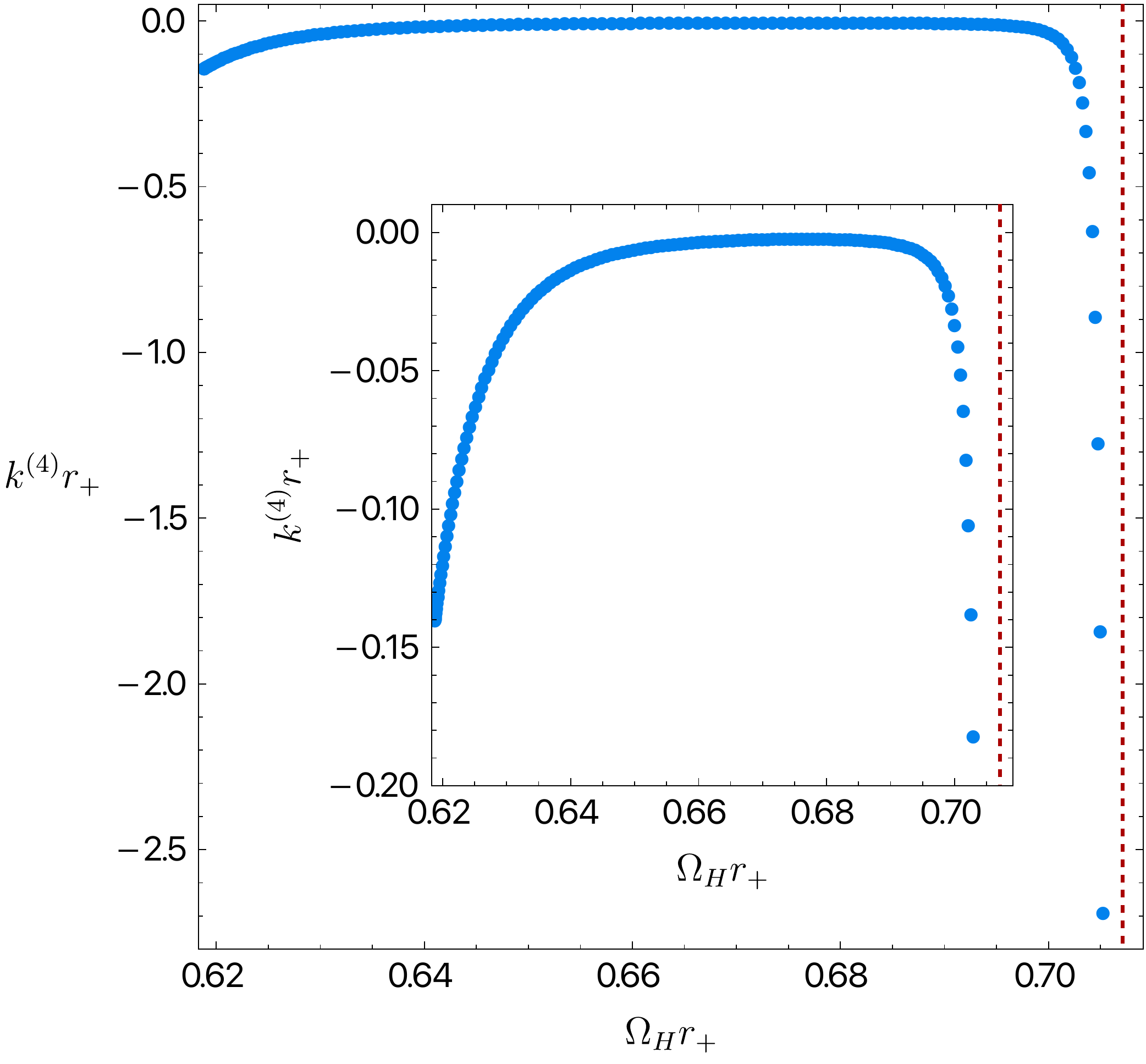}
\caption{Wavenumber corrections $\widetilde{k}^{(2)}$ (left panel) and $\widetilde{k}^{(4)}$ (right panel), as defined in \eqref{PTexpansionB}, as a function of the dimensionless angular velocity $\Omega r_+ =\widetilde{a}$. The vertical red dashed line represents the extremal configuration with $\Omega_H r_+=1/\sqrt{2}$. Recall that the leading wavenumber $\widetilde{k}^{(0)}$ is given in Fig.~\ref{Fig:zeroModeSuperGL}.
} \label{Fig:k2k4}
\end{figure}

As we expand to higher orders, we find that the odd and even orders in perturbation theory behave differently. Even orders $\mathcal{O}(\epsilon^n)$ introduce perturbative corrections to thermodynamic quantities like  energy, angular momenta, tension, entropy, temperature, angular velocity, etc, but they do not correct the wavenumber, $\widetilde{k}^{(n-1)}=0$ (and thus do not correct $\widetilde{L}$).

Odd orders $\mathcal{O}(\epsilon^n)$ give the wavenumber corrections $\widetilde{k}^{(n-1)}$ but do not change the thermodynamic quantities. The leading wavenumber $\widetilde{k}_{(0)}$ was already shown in Fig.~\ref{Fig:zeroModeSuperGL}. On the other hand, the next-to-leading order wavenumber corrections $\widetilde{k}^{(2)}$  and $\widetilde{k}^{(4)}$, as defined in \eqref{PTexpansionB}, are plotted in Fig.~\ref{Fig:k2k4}, in the left and right panels, respectively. The fact that (some of) these higher order quantities grow large as one approaches $\widetilde{\Omega}_H=\widetilde{\Omega}_H|_c$ and/or $\widetilde{\Omega}_H=\widetilde{\Omega}_H^{\rm ext}=1/\sqrt{2}$ tells us that our perturbation theory breaks down in this region. We will come back to this below.

Once we have found all the Fourier coefficients $\mathfrak{q}_j^{(n,\eta)}(y)$ and wavenumber corrections $\widetilde{k}^{(n-1)}$, we can reconstruct the eight fields $q_j(x,y)$ using \eqref{PTexpansion}. We can then substitute these fields in the thermodynamic formulas \eqref{ThermoResonatorDless} of section~\ref{sec:thermo} to obtain all the thermodynamic quantities of the system up to $\mathcal{O}(\epsilon^5)$.
We complete this perturbation scheme up to order $\mathcal{O}(\epsilon^5)$: this is the order required to find a deviation between the relevant thermodynamics of the black resonator and MP strings, as it will be found when obtaining \eqref{EntropyDiff}.

Having described the perturbation scheme, we are now ready to discuss the system's properties that can be extracted from the perturbative analysis. First of all, let us recall that it follows from \eqref{ThermoResonatorDless} (with $\alpha^{(1)}_{1,6,7}=0$ and $q_{4,5,7}(x,0)=1$) that the thermodynamic quantities of  MP strings parametrized by $(\widetilde{L},\widetilde{a})$ are given by
\begin{eqnarray}\label{ThermoMPonset}
&&\mathcal{E}\big|_{\hbox{\tiny MP}}=\frac{1}{G_6}\frac{3 \pi}{8 \widetilde{L}^2} \frac{1}{1-\widetilde{a} ^2}\,, \qquad \mathcal{J}\big|_{\hbox{\tiny MP}}=\frac{1}{G_6}\frac{\pi}{4 \widetilde{L}^3}\, \frac{\widetilde{a}}{1-\widetilde{a} ^2}\,, \qquad \mathcal{T}_z\big|_{\hbox{\tiny MP}}=\frac{1}{G_6}\frac{\pi}{8 \widetilde{L}^2}\,\frac{1}{1-\widetilde{a} ^2},
\nonumber \\
&&\tau_H\big|_{\hbox{\tiny MP}}=\frac{\widetilde{L}}{2 \pi} \frac{1-2 \widetilde{a} ^2}{\sqrt{1-\widetilde{a} ^2}}\,, \qquad \omega_H\big|_{\hbox{\tiny MP}}= \widetilde{a} \,\widetilde{L}\,, \qquad \sigma_H\big|_{\hbox{\tiny MP}}=\frac{1}{G_6} \frac{\pi^2}{2 \widetilde{L}^3}\frac{1}{\sqrt{1-\widetilde{a} ^2}} \,.
\end{eqnarray}
Extremal MP strings (i.e. with $\tau_H=0$) are a 1-parameter family of solutions with $\widetilde{\Omega}_H=\widetilde{\Omega}_H^{\rm ext}=1/\sqrt{2}$ and
\begin{equation}\label{MPext:EJ}
\mathcal{J}_{\hbox{\tiny ext\, MP}}(\mathcal{E})=\frac{2^{3/2}}{3^{3/2}\pi^{1/2}}G_6^{1/6}\mathcal{E}^{3/2}\;.
\end{equation}

\begin{figure}[t!]
\centerline{
\includegraphics[width=.70\textwidth]{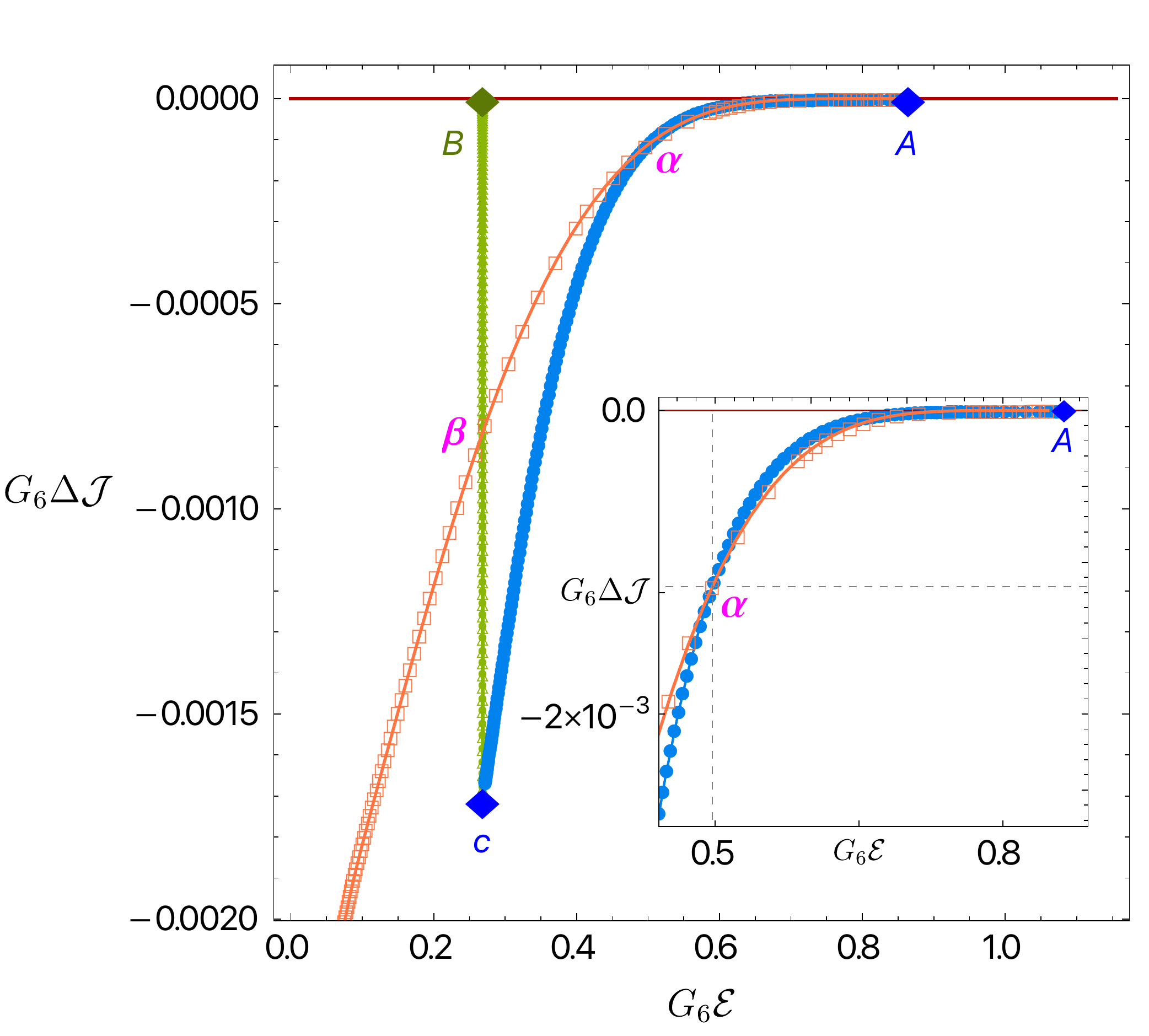}
}
\caption{Superradiant instability (for $m=2$) and Gregory-Laflamme instability of Myers-Perry black strings with parameters $\mathcal E$, $\mathcal J$.  For presentation, we show the momentum difference with the extremal Myers-Perry black string $G_6 \Delta \mathcal{J}\equiv G_6(\mathcal{J}-\mathcal{J}_{\hbox{\tiny ext\, MP}})_{\hbox{\tiny same}\,\mathcal{E}}$. Superradiant instability occurs inside the triangular region $ABc$, and Gregory-Laflamme instability occurs above the curve marked by orange squares. The Gregory-Laflamme onset curve intersects with the edge of the unstable superradiant region at points $\alpha$ and $\beta$. The horizontal red line at $\Delta \mathcal{J}=0$ is extremality.
}
\label{fig:stabilityDiag}
\end{figure}

At leading order in perturbation theory, we can recover our results from \cite{Dias:2022mde}, which was already presented in Fig.~\ref{Fig:zeroModeSuperGL}.  Here, in  Fig.~\ref{fig:stabilityDiag}, we show the results again but this time in terms of the dimensionless energy and angular momentum (in units of circle length $L$) as defined in section~\ref{sec:thermo}. For visibility, it is convenient to show the angular momentum difference from the extremal Myers-Perry black string
\begin{equation}\label{DeltaJ}
\Delta \mathcal{J}\equiv (\mathcal{J}-\mathcal{J}_{\hbox{\tiny ext\, MP}})|_{\hbox{\tiny same}\,\mathcal{E}},
\end{equation}
rather than $\mathcal J$ itself. 

The various quantities that are plotted in Fig.~\ref{fig:stabilityDiag} are the same as that of Fig.~\ref{Fig:zeroModeSuperGL}.  The triangular region $ABc$ is where Myers-Perry black strings are unstable to superradiance. In particular, the blue curve $Ac$ represents the onset curve of the superradiant instability.

This onset curve $Ac$ is where a new family of black resonator strings branches from Myers-Perry black strings. Perturbation theory at order $\mathcal{O}(\epsilon)$ locates this curve, but it cannot describe the thermodynamic properties of the black resonator strings.  For that, we need to proceed to higher order.  We do so up to $n=5$ and thus we get the thermodynamic description of black resonator strings up to $\mathcal{O}(\epsilon^5)$. We can then compare it against the thermodynamics of MP strings and find which of these two families is the preferred phase, when they coexist. We are particularly interested in the microcanonical ensemble, so the dominant  phase is the one that has the highest  dimensionless entropy $\sigma_H$ for given energy $\mathcal{E}$ and angular momenta $\mathcal{J}$ (in units of circle length $L$)

To complete this task, let $\mathcal{Q}_{\rm MP}$ and $\mathcal{Q}_{\rm res}$ denote generic thermodynamic quantities $\mathcal{Q}$ for the Myers-Perry black string and black resonator string, respectively. 
Again, when comparing these two solutions in the microcanonical ensemble, we must use the same Kaluza-Klein circle size $L$.  Accordingly, we must work with quantities in units of $L$ \eqref{dimensionlessThermo}.  Then, in these units, we also require the solutions to have the same dimensionless energy and angular momentum:
\begin{equation}\label{microCondition}
\mathcal{E}_{\rm res}=\mathcal{E}_{\hbox{\tiny MP}}\,, \qquad \mathcal{J}_{\rm res}=\mathcal{J}_{\hbox{\tiny MP}}.
\end{equation}
Given a resonator string with $(\mathcal{E}_{\rm res},\mathcal{J}_{\rm res})$ we must therefore identify the corresponding Myers-Perry black string with the same energy and angular momentum.  To translate to the parameters $(\widetilde{L}_{\hbox{\tiny MP}},\widetilde{a}_{\hbox{\tiny MP}})$ that we have used to describe the Myers-Perry black strings, we use the following relations: 

\begin{align}
& \widetilde{L}_{\hbox{\tiny MP}} =\sqrt{3 \pi} \, \frac{\sqrt{\mathcal{E}_{\rm res}}}{\Sigma_+} \,,
  \qquad
 \widetilde{a}_{\hbox{\tiny MP}}= \frac{1}{12 \sqrt{3 \pi}}\, \frac{\Sigma_+ \Sigma_-^2}{\mathcal{J}_{\rm res} \mathcal{E}_{\rm res}^{3/2}} \nonumber \\
& \hbox{with} \quad  \Sigma_\pm \equiv \sqrt{4\mathcal{E}_{\rm res}^2 \pm  \sqrt{2}\sqrt{ 8\mathcal{E}_{\rm res}^4-27 \pi  \mathcal{J}_{\rm res}^2 \mathcal{E}_{\rm res}}}
\end{align}
We can now replace these quantities in \eqref{ThermoMPonset} with the identifications $ \widetilde{L}\to \widetilde{L}_{\hbox{\tiny MP}}$ and   $ \widetilde{a}\to \widetilde{a}_{\hbox{\tiny MP}}$ 
to find the thermodynamic quantities (in particular, the entropy $\sigma_H$) of the Myers-Perry string with the same energy and angular momenta as the resonator string.

This process gives us a family of Myers-Perry black strings in terms of $(\mathcal{E}_{\rm res},\mathcal{J}_{\rm res})$.  Since, in our perturbation scheme, the resonator quantities depend on $\epsilon$, so too do the corresponding Myers-Perry black strings. We can now compute the entropy density difference $\Delta \sigma_H=\left( \sigma_{H,\rm res}-\sigma_{H,\hbox{\tiny MP}}\right)|_{{\rm same}\,(\mathcal{E},\mathcal{J})}$ between resonator strings and Myers-Perry strings. This gives $\Delta \sigma_H=c_{\Delta \sigma}^{(2)}\epsilon^2+ c_{\Delta \sigma}^{(4)}\epsilon^4$ where $c_{\Delta \sigma}^{(2)}$ and $c_{\Delta \sigma}^{(4)}$ are two functions of $\mathfrak{q}_j^{(n,\eta)}|_{y=0}$ ($n=2,4$). 

We can use the first law of thermodynamics \eqref{1stLawDless} to show that $c_{\Delta \sigma}^{(2)}= 0$ which justifies our need to extend the perturbation expansion up to $\mathcal{O}(\epsilon^5)$. We also find that $c_{\Delta \sigma}^{(4)}$ no longer depends on order $n=4$ functions $\mathfrak{q}_j^{(4,\eta)}$.\footnote{Note that the energy, etc are even functions of $\epsilon$ and the first law \eqref{1stLawDless} can be written as $\partial_\epsilon \mathcal{E}=  \tau_H\,\partial_\epsilon\sigma_H+ 2\,\omega_H \, \partial_\epsilon\mathcal{J}$. The first law must be obeyed at each order in $\epsilon$ and thus it effectively gives two conditions (one at order $\epsilon$ and the other at order $\epsilon^3$) that we can use to express the second derivatives of $\mathfrak{q}_7^{(2,0)}|_{y=1}$ and $\mathfrak{q}_7^{(4,0)}|_{y=1}$ as a function of other functions $\mathfrak{q}_j^{(n,\eta)}$ and their first derivatives evaluated at the horizon, $y=0$, or at $y=1$. When we do this, we simplify considerably $c_{\Delta \sigma}^{(2)}$ and $c_{\Delta \sigma}^{(4)}$. In particular, this finds that $c_{\Delta \sigma}^{(2)}=0$. As a further check of our numerics, we verify that the Smarr law \eqref{SmarrDless} is obeyed by our solutions.} Altogether, after using the first law of \eqref{1stLawDless}, we find that

\begin{eqnarray} \label{EntropyDiff}
\Delta \sigma_H&=&\left( \sigma_{H,\rm res}-\sigma_{H,\hbox{\tiny MP}}\right)\big|_{{\rm same}\,(L,\mathcal{E},\mathcal{J})}\nonumber \\
&=&c_{\Delta \sigma}^{(4)}\,\epsilon^4  +\mathcal{O}(\epsilon^6)
\end{eqnarray}
with\footnote{The entropy difference depends also on functions evaluated at the asymptotic boundary because we have subtracted the Myers-Perry background
and because we used first law to get \eqref{EntropyDiff2}.}
\begin{eqnarray} \label{EntropyDiff2}
c_{\Delta \sigma}^{(4)}& =& \frac{\widetilde{k}_{(0)}^2}{384 \pi  \sqrt{1-\widetilde{a} ^2} \left(1-2 \widetilde{a} ^2\right)}\Bigg\{  \widetilde{k}_{(0)} \Bigg[
2 \left(1-\widetilde{a} ^2\right)^2 \left(3-2 \widetilde{a} ^2\right) \widetilde{a} ^2  \mathfrak{q}_6^{(2,0)\,\prime}(1)^2
\nonumber \\
& &
-8 \widetilde{a} ^2 \left(\widetilde{a} ^4-3 \widetilde{a} ^2+2\right)  \Bigg(\mathfrak{q}_4^{(2,0)}(0)+2 \mathfrak{q}_5^{(2,0)}(0)+\mathfrak{q}_7^{(2,0)}(0)\Bigg) \mathfrak{q}_6^{(2,0)\,\prime}(1) \nonumber \\
& &
 -\left(4 \widetilde{a} ^4-10 \widetilde{a} ^2+1\right) \Big(\mathfrak{q}_4^{(2,0)}(0)+2 \mathfrak{q}_5^{(2,0)}(0)+\mathfrak{q}_7^{(2,0)}(0)\Big)^2\Bigg] \\
& &
-6 \widetilde{k}^{(2)} \Bigg[\left(1-2 \widetilde{a} ^2\right) \Bigg(\mathfrak{q}_4^{(2,0)}(0)+2 \mathfrak{q}_5^{(2,0)}(0)+\mathfrak{q}_7^{(2,0)}(0)\Bigg)+2 \left(1-\widetilde{a} ^2\right) \widetilde{a} ^2  \mathfrak{q}_6^{(2,0)\,\prime}(1)\Bigg]
\Bigg\},\nonumber
\end{eqnarray}
where  $\mathfrak{q}_6^{(2,0)\,\prime}(1)$ stands for the first derivative of $\mathfrak{q}_6^{(2,0)}$ evaluated at $y=1$ and all other functions in \eqref{EntropyDiff2} are evaluated at $y=0$.

\begin{figure}[th]
\centering
\includegraphics[width=.6\textwidth]{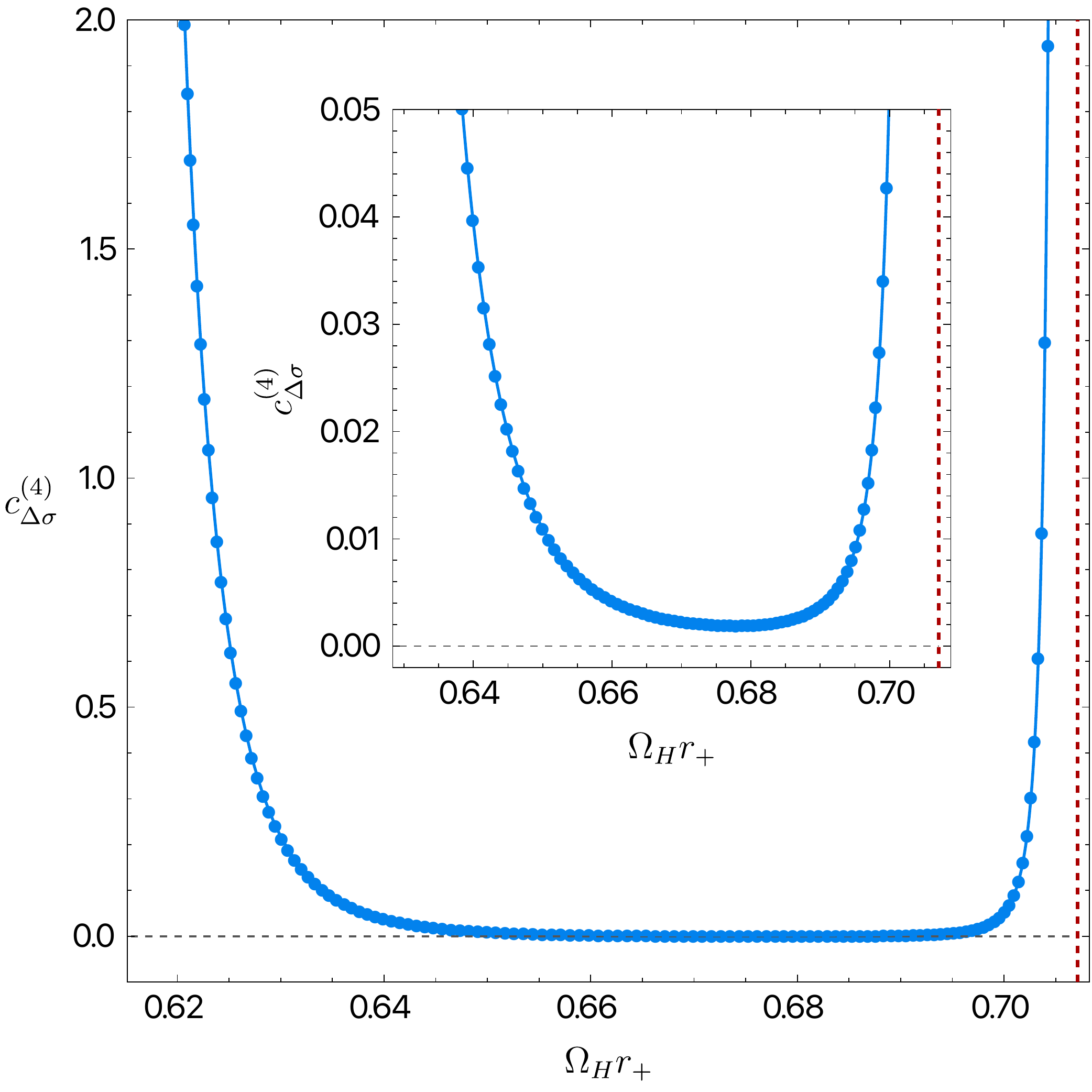}
\caption{Lowest-order perturbative correction to the entropy of Myers-Perry black strings by branching $m=2$ black resonator strings near the onset of the superradiant instability. The horizontal axis shows the dimensionless angular velocity $\widetilde{\Omega}_H$ of the strings. The vertical axis shows the difference \eqref{EntropyDiff2} between the dimensionless entropy of the resonator and Myers-Perry black strings with the same $(L,\mathcal{E},\mathcal{J})$. Since $c_{\Delta \sigma}^{(4)}>0$, black resonator strings always dominate the microcanonical ensemble around the superradiant merger line.  The divergences near endpoints imply that perturbation theory is breaking down at these locations. (See also later Fig.~\ref{fig:nearonset}).
}
\label{Fig:DeltaSigma}
\end{figure}

For a given $(\mathcal{E},\mathcal{J})$, if $\Delta\sigma_H>0$ then the black resonator strings are the preferred phase, at least in a neighbourhood of the $m=2$ superradiant onset. Recall that this onset is given by the curve $Ac$ in Fig.~\ref{Fig:zeroModeSuperGL} or Fig.~\ref{fig:stabilityDiag} in the range $\widetilde{\Omega}_H|_c \leq \widetilde{\Omega}_H \leq \widetilde{\Omega}_H^{\rm ext}$. Thus, to analyse the positivity of $\Delta\sigma_H$, we  just need to compute the coefficient  $c_{\Delta \sigma}^{(4)}$ in \eqref{EntropyDiff}-\eqref{EntropyDiff2} as a function of $ \widetilde{\Omega}_H $. This is done in Fig.~\ref{Fig:DeltaSigma}. We conclude that, for any value  of $\widetilde{\Omega}_H|_c \leq \widetilde{\Omega}_H \leq \widetilde{\Omega}_H^{\rm ext}$,  $c_{\Delta \sigma}^{(4)}$ and thus $\Delta\sigma_H$ are positive quantities. It follows that black resonator strings near the superradiant onset have a higher entropy than the corresponding Myers-Perry black strings.

From the perturbative analysis, we find that the black resonator strings branch from the Myers-Perry black strings in a direction towards the unstable region  $Ac$ of  Fig.~\ref{fig:stabilityDiag}  (see later Fig.~\ref{fig:nearonset}).\footnote{Note that as explained above, $\Delta \sigma_H$,  $\mathcal{E}$ and $\mathcal{J}$ have an expansion in $\epsilon$ and thus we can do a parametric plot of $\Delta \sigma_H$ as a function of $\mathcal{E}$ and $\mathcal{J}$.}  The fact that the entropy has an expansion in even powers of $\epsilon$ also implies that perturbations with different signs are equivalent.  This means that the black resonator strings only extend in one direction\footnote{Black hole resonators that branch from the onset of other superradiant systems \cite{Dias:2011at,Dias:2011tj,Dias:2015rxy,Ishii:2018oms,Ishii:2020muv,Ishii:2021xmn,Herdeiro:2014goa,Dias:2018yey} have this property. The non-uniform strings that branch from the Gregory-Laflamme onset also have this property \cite{Gubser:2001ac,Harmark:2002tr,Kol:2002xz,Wiseman:2002zc,Kol:2003ja,Harmark:2003dg,Harmark:2003yz,Kudoh:2003ki,Sorkin:2004qq,Gorbonos:2004uc,Kudoh:2004hs,Dias:2007hg,Harmark:2007md,Wiseman:2011by,Figueras:2012xj,Dias:2017coo}. However, there are also systems where the changing the sign of the perturbation gives different physical results, and these have solutions that branch in two directions from onset of the instability \cite{Dias:2014cia,Emparan:2014pra,Dias:2015nua,Dias:2009iu,Dias:2010maa,Dias:2010eu,Dias:2010maa,Dias:2010gk,Dias:2011jg,Dias:2017coo,Dias:2015pda,Dias:2016eto}.}.  Together with the entropy results, this implies that it is entropically permissible for unstable Myers-Perry black strings near the onset to evolve to black resonator strings.

This perturbative analysis, however, has its limitations.  First, we note from Fig.~\ref{Fig:DeltaSigma} that the entropy correction $c_{\Delta \sigma}^{(4)}$ is diverging near the endpoints of the merger line, which is an indication that perturbation theory is breaking down in such regions. Second, the perturbative analysis does not say anything about solutions that are far from the merger line.  We will construct such solutions by solving the full Einstein equation numerically and study their thermodynamics in section~\ref{sec:PhaseDiag}. We will make a direct comparison between numerics and perturbation theory later in Fig.~\ref{fig:nearonset}.

\section{Constructing geons with Kaluza-Klein asymptotics \label{sec:geons}}

Besides the existence of black resonators we will also find another intriguing class of solutions. These are horizonless configurations that are smooth and are analogous to AdS geons detailed in \cite{Dias:2011ss,Horowitz:2014hja,Martinon:2017uyo,Ishii:2018oms}. For this reason we will refer to these as {\it Kaluza-Klein geons}. To find these solutions we again use the DeTurck method \cite{Headrick:2009pv,Figueras:2011va,Wiseman:2011by,Dias:2015nua,Figueras:2016nmo}. We start with presenting a line element consistent with our symmetries, just like we did in \eqref{ansatz} for the black string resonators, but now without an horizon. We take
\begin{multline}
\mathrm{d}s^2 = -Q_1\,\mathrm{d}T^2+\frac{4\,Q_2\,\mathrm{d}y^2}{(1-y^2)^4(2-y^2)}+Q_7\,\left(\frac{\widetilde{L}}{2}\mathrm{d}x+Q_3\,\mathrm{d}y\right)^2
\\
+\frac{y^2(2-y^2)}{(1-y^2)^2}\left\{Q_4\,\left[\frac{\Sigma_3}{2}+\Omega_0\,\mathrm{d}T+\Omega_0(1-y^2)^3Q_6\, \mathrm{d}T\right]^2+\frac{Q_5}{4}\left(Q_8\,\Sigma_1^2+\frac{\Sigma_2^2}{Q_8}\right)\right\}\,,
\label{eq:ansatz_geons}
\end{multline}
where $Q_i(x,y)$, with $i\in\{1,\ldots,8\}$ are functions of $x$ and $y$ only and $\widetilde{L}$ and $\Omega_0$ are constants, whose physical meaning we will discuss shortly.

For the reference metric in the De Turck method we choose
\begin{equation}
Q_1=Q_2=Q_4=Q_5=Q_7=Q_8=1\,\quad\text{and}\quad Q_6=Q_3=0\,,
\end{equation}
which is just the Kaluza-Klein spacetime $\mathcal{M}^{1,4}\times S^1$.

Note that in \eqref{eq:ansatz_geons} $y=0$ is a regular smooth center and (for any constant $x$ and $T$), it corresponds to a point in spacetime. Spatial infinity is located at $y=1$, and here $x$ is a periodic coordinate with period $x\sim x+2$. Without loss of generality, we take $x\in[-1,1]$. We are further interested in solutions which preserve a $\mathbb{Z}_2$ symmetry around $x=0$ and are invariant under $x\to-x$. For this reason we will take $x\in[0,1]$ and choose boundary conditions at $x=0$ and $x=1$ consistent with this discrete symmetry. This amounts to choosing
\begin{align}
&\left.\frac{\partial Q_i}{\partial x}\right|_{x=0}=Q_3(0,y)=0\,,\quad \forall_{i\neq3}\,,\nonumber
\\
&\left.\frac{\partial Q_i}{\partial x}\right|_{x=1}=Q_3(1,y)=0\,,\quad \forall_{i\neq3}\,.\nonumber
\end{align}

Our integration domain is thus a unit square $(0,1)\times(0,1)$. At spatial infinity we demand the line element to approach the reference metric, and at $y=0$ we demand regularity. This last statement in turn implies
\begin{equation}
\left.\frac{\partial Q_i}{\partial y}\right|_{y=0}=Q_3(x,0)=0\,,\quad \forall_{i\neq3}.
\end{equation}

We are left with discussing the physical meaning of $\widetilde{L}$ and $\Omega_0$. In the absence of a black hole, which necessarily introduces a novel scale, there is a special symmetry of the Einstein DeTurck equations for the ansatz shown in \eqref{eq:ansatz_geons}. In particular, it is a simple exercise to show that under the scaling
\begin{equation}
Q_1= \frac{\tilde{Q}_1}{\widetilde{L}^2}\,,\quad Q_7= \frac{\tilde{Q}_7}{\widetilde{L}^2}\,,\quad Q_3=\widetilde{L}\,\tilde{Q}_3\quad\text{and}\quad Q_i=\tilde{Q}_i\quad \forall_{i\neq1,7,4}\,,
\end{equation}
the Einstein DeTurck equations for the $\tilde{Q}$ depend only on the product $\widetilde{L}\,\Omega_0$. This scaling shows that, despite depending on two parameters, Eq.~(\ref{eq:ansatz_geons}) really only yields a one-parameter family of physical solutions. One can either decide to fix $L$, and change $\Omega_0$ to move along the family of solutions, or vice-versa. Here $L$ can be interpreted as the length of the Kaluza-Klein circle at spatial infinity, and $\Omega_0$ as the geon angular velocity.

One can apply the procedure outlined in section \ref{sec:thermo} to compute the thermodynamic properties of the Kaluza-Klein geon. We note that the metric \eqref{eq:ansatz_geons} is written in a frame that is rotating at infinity. To accommodate this, we first change to a different angular coordinate $\Psi = \psi-\Omega_0 t$ and only afterwards compute all the relevant thermodynamic quantities. This in turn implies that the geon angular velocity is give by $\omega_H=\Omega_0\,\widetilde{L}$. Note that though we continue to use the subscript $H$, the dimensionless angular velocity $\omega_H$ really measures the angular velocity of the Kaluza-Klein geon with respect to a static observer at infinity, as there is no horizon for geons.

We defer the discussion of our results on this calculation to section \ref{sec:PhaseDiag}.


\section{No warm holes with Kaluza-Klein asymptotics \label{sec:warmholes}}
There is yet another class of geometries that could potentially play an important role in our discussion. These are black hole solutions that are on the verge of becoming scattering states, and yet manage to have finite energy, entropy and temperature. They have been first uncovered in \cite{Dias:2021vve} in the context of four-dimensional asymptotically flat charged black holes coupled to a charged scalar field with non-minimal couplings and were coined \emph{warm holes}.

In the present context, Kaluza-Klein warm holes, if they exist, would be black resonator strings that would have the maximum or minimum angular momenta (at least for a given window of lengths and energies) and they should have finite entropy and nonzero temperature.  If we lower the temperature of the system towards zero, it is possible for solutions to stop existing at a nonzero temperature because such solutions can no longer be confined by the gravitational potential created by the Kaluza-Klein momentum. Such a minimum, non-zero temperature black string would be called a warm hole (or, if we prefer, warm string) if and only if it happens to be a smooth solution. 

One way to understand the status of warm hole solutions for our ansatz comes from solving the equations of motion for $q_8$ appearing in \eqref{ansatz} asymptotically. By performing a generalised Frobenius analysis near asymptotic infinity, one finds
\begin{subequations}
\label{eqs:enco}
\begin{equation}
q_8(x,y)\approx 1+e^{-\frac{\hat{\lambda}}{1-y}}\cos(\pi x)(1-y)^{\frac{3}{2}}\left\{a_0 +\mathcal{O}[(1-y)]\right\}+\mathcal{O}\left[e^{-\frac{2\hat{\lambda}}{1-y}}\right]\,,
\end{equation}
with
\begin{equation}
\hat{\lambda}\equiv \sqrt{\frac{\pi ^2}{\widetilde{L}^2}-4 \widetilde{a} ^2}\,.
\end{equation}
\end{subequations}%

Warm hole solutions are marginally bound (i.e. on the verge of not being confined by the potential created by the Kaluza-Klein momentum), for which the exponential decay in $q_8$ near $y=1$ is absent, \emph{i.e.}
\begin{equation}
\widetilde{L}=\frac{\pi}{2|\widetilde{a}|}\,.
\label{eq:Ltilde}
\end{equation}
In this case, the expansion encoded in \eqref{eqs:enco} is no longer valid. Instead, one finds two possibilities
\begin{equation}
q_8(x,y)-1\approx \cos(\pi x)(1-y)^{1\pm\sqrt{9-16 \pi G_6(\mathcal{E}-\mathcal{T}_z)}}\left[1+\mathcal{O}(1-y)\right]\,.
\label{eq:exp}
\end{equation}

A few comments are in order regarding the above expansion. One can show that the procedure for computing all the relevant thermodynamics of the solution remains unchanged from what we saw in section \ref{sec:thermo}, so long as we take the upper sign in \eqref{eq:exp} and the argument of the square root is positive definite. This means that we can take all the relevant thermodynamic quantities in said section, and replace $\widetilde{L}$ as given in \eqref{eq:Ltilde}. If $9-16 \pi G_6(\mathcal{E}-\mathcal{T}_z)>0$, the upper sign gives rise to normalisable solutions, whereas the minus sign in \eqref{eq:exp} leads to unphysical solutions with infinite energy. Numerically, we find that $9-16 \pi G_6(\mathcal{E}-\mathcal{T}_z)<0$ throughout parameter space, in particular in the limit where $\widetilde{L}$ approaches $\pi/(2\widetilde{a})$ from below. This rules out the existence of smooth, normalisable warm holes solutions with Kaluza-Klein asymptotics that connect smoothly to our black resonator strings.

As we shall see in Section~\ref{sec:PhaseDiag}, there is a limit of black resonator strings where they fail to become bound states.  What we have shown in this section is that such limiting solutions cannot be regular.

\section{Phase diagram of black resonator strings \label{sec:PhaseDiag}}

\begin{figure}[th]
\centering
\includegraphics[width=.5\textwidth]{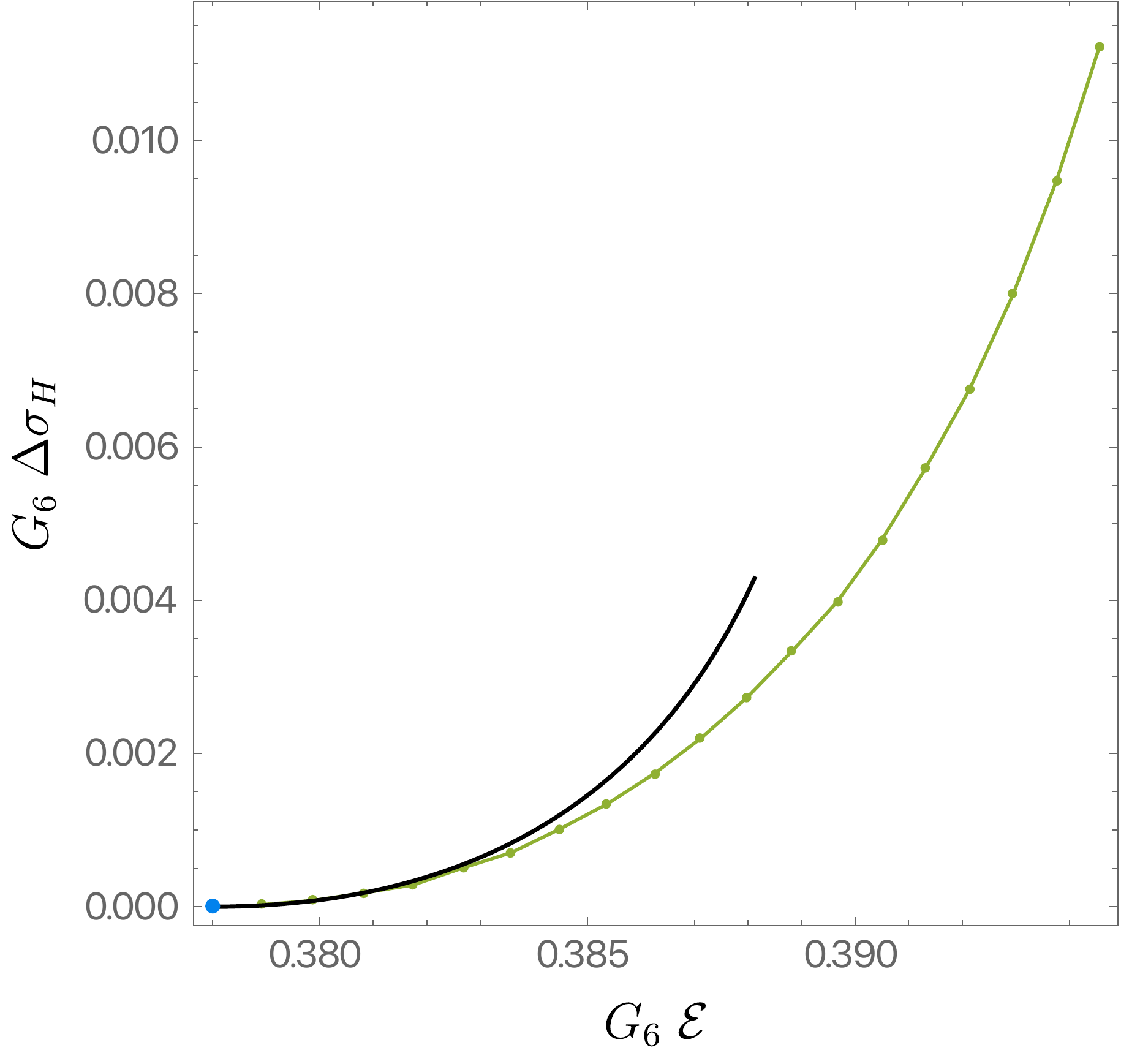}
\caption{Comparison between nonlinear solutions and the perturbative results of section \ref{sec:perturbative}.  The solid black line is the perturbative expansion, whereas the green disks are the full nonlinear results. Both curves were generated with $\widetilde{\Omega}_H=0.66$ (though recall that $\Delta\sigma_H$ is the difference with respect to the Myers-Perry black string at the same $\mathcal E$ and $\mathcal J$, so the reference Myers-Perry black string may have a different $\widetilde{\Omega}_H$).  The Myers-Perry superradiant onset is the blue disk with $\Delta\sigma_H=0$. The fact that $\Delta \sigma_H\geq0$ indicates that black resonators are the dominant configuration.}
\label{fig:nearonset}
\end{figure}

In this section we discuss the phase space of solutions among black resonator strings, Kaluza-Klein geons, and Myers-Perry black strings. We find that whenever solutions co-exist with the same $\mathcal{E}$ and $\mathcal{J}$, the black resonator strings have the highest entropy. In Fig.~\ref{fig:nearonset}, we show the entropy difference between a black resonator string and a Myers-Perry black string (at the same $\mathcal{E}$ and $\mathcal{J}$ as in \eqref{EntropyDiff}), for black resonator strings at constant $\widetilde{\Omega}_H=0.66$. The black solid line shows the perturbative result of \eqref{EntropyDiff}-\eqref{EntropyDiff2} constructed in section~\ref{sec:perturbative}, whereas the green disks give the fully nonlinear numerical data. The agreement between the two methods near the onset is reassuring.

\begin{figure}[th]
\centering
\includegraphics[width=.7\textwidth]{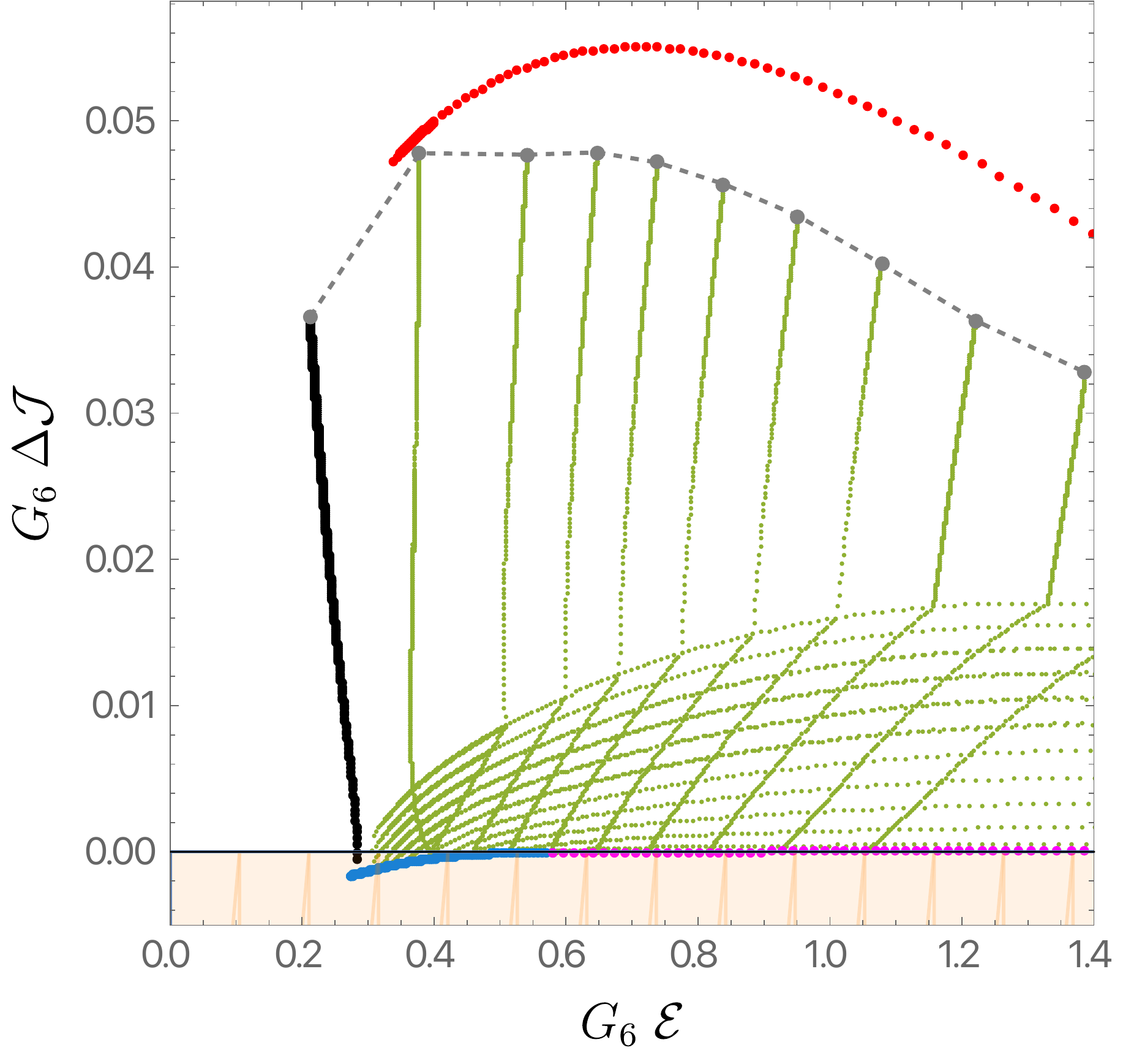}
\caption{Phase space of black resonator string solutions and Kaluza-Klein geons. The green points are black resonator strings and the red points are Kaluza-Klein geons. As far as we can tell, black resonator strings and Kaluz-Klein geons do not overlap.  Black resonator strings merge with Myers-Perry black strings at the onset curve given by blue points.  The magenta points are our best estimate of where black resonator string are extremal. The black points on the left are where black resonator strings cease being bound states.  The gray points mark turning points where black resonator strings have maximum angular momenta.  There is (at least) a second set of black resonator strings that exist just below this gray curve, ending in a curvature singularity (which also lies somewhere just below this gray curve).}
\label{Fig:phaseDiagResonator}
\end{figure}

We now proceed to show the full phase space of solutions in Fig.~\ref{Fig:phaseDiagResonator}.  For presentation, we show solutions parametrised by $(\mathcal{E},\Delta \mathcal{J})$, where $\Delta \mathcal{J}$ gives the difference in angular momentum between a given solution and the corresponding \emph{extremal} Myers-Perry black string with the same energy $\mathcal{E}$, \emph{i.e.} $\Delta \mathcal{J}=\mathcal{J}-\mathcal{J}_{\hbox{\tiny ext\, MP}}$. 

\begin{figure}[th]
\centering
\includegraphics[width=.5\textwidth]{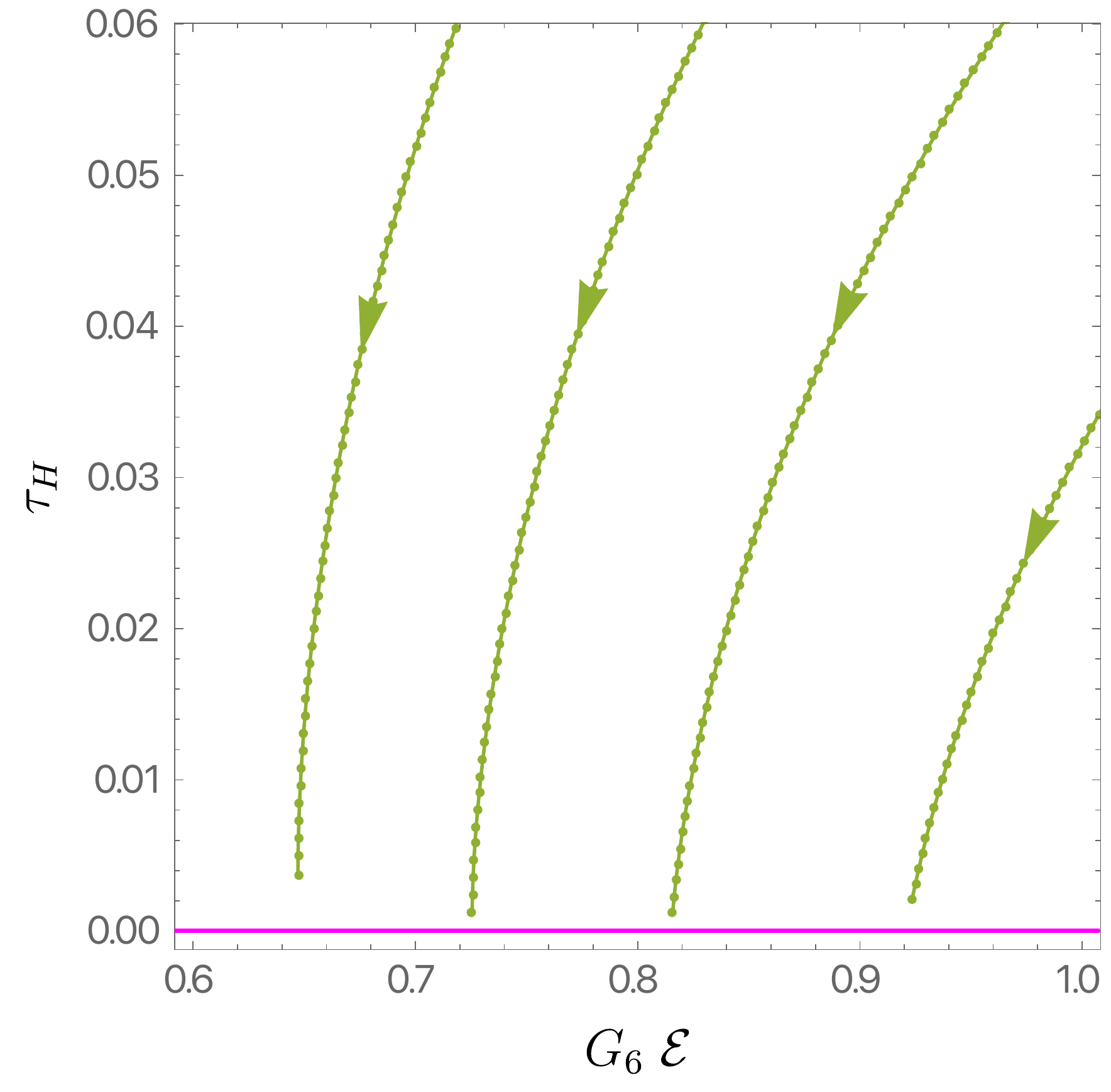}
\caption{$\tau_H$ as a function of $\mathcal{E}$ for several of the solutions in Fig.~\ref{Fig:phaseDiagResonator} that approach extremality. The direction of the arrows indicates decreasing values of $\Delta \mathcal{J}$ in Fig.~\ref{Fig:phaseDiagResonator}.}
\label{fig:temperature}
\end{figure}
Black resonator strings also have their own extremal limit, which we estimate to be the magenta line in Fig.~\ref{Fig:phaseDiagResonator}. To illustrate this, we show in Fig.~\ref{fig:temperature} the temperature $\tau_H$ as a function of $\mathcal{E}$ of four lines of black resonator strings. The arrows in the lines of Fig.~\ref{fig:temperature}  indicate the direction of decreasing values of $\Delta \mathcal{J}$.  By approaching as close as possible to zero temperature across many such curves, we obtain the set of magenta points displayed in Fig.~\ref{Fig:phaseDiagResonator}.

The black disks in Fig.~\ref{Fig:phaseDiagResonator} are the lowest-energy black resonator strings we have managed to construct.  Black resonator strings do not exist too far to the left of this curve because they are no longer bound states.  The lowest-energy limit solutions of black resonator strings here would be warm holes, but as we have shown in \eqref{eq:exp}, warm holes are not regular solutions.  We emphasize that constructing solutions near the black disks is a numerical challenge, so we do not have precise knowledge of where the lowest-energy black resonator strings lie in phase space.  In particular, we do not know if this low-energy limit lies to the right, left, or intersects the confining cutoff curve for $m=2$ perturbations of Myers-Perry black strings (shown in green in Fig.~\ref{fig:stabilityDiag}).  Unfortunately, this means that we do not know if the phase space of black resonator strings completely covers the region of unstable (to $m=2$ superradiant perturbations) Myers-Perry black holes or not.

The red points in Fig.~\ref{Fig:phaseDiagResonator} are the Kaluza-Klein geons of section~\ref{sec:geons} and, as far as we can tell, they neither coexist nor are a limiting solution of black resonator strings. In other words, black resonator strings and the Kaluza-Klein geons are disconnected in the phase diagram and seem to be unrelated solutions.

\begin{figure}[th]
\centering
\includegraphics[width=.5\textwidth]{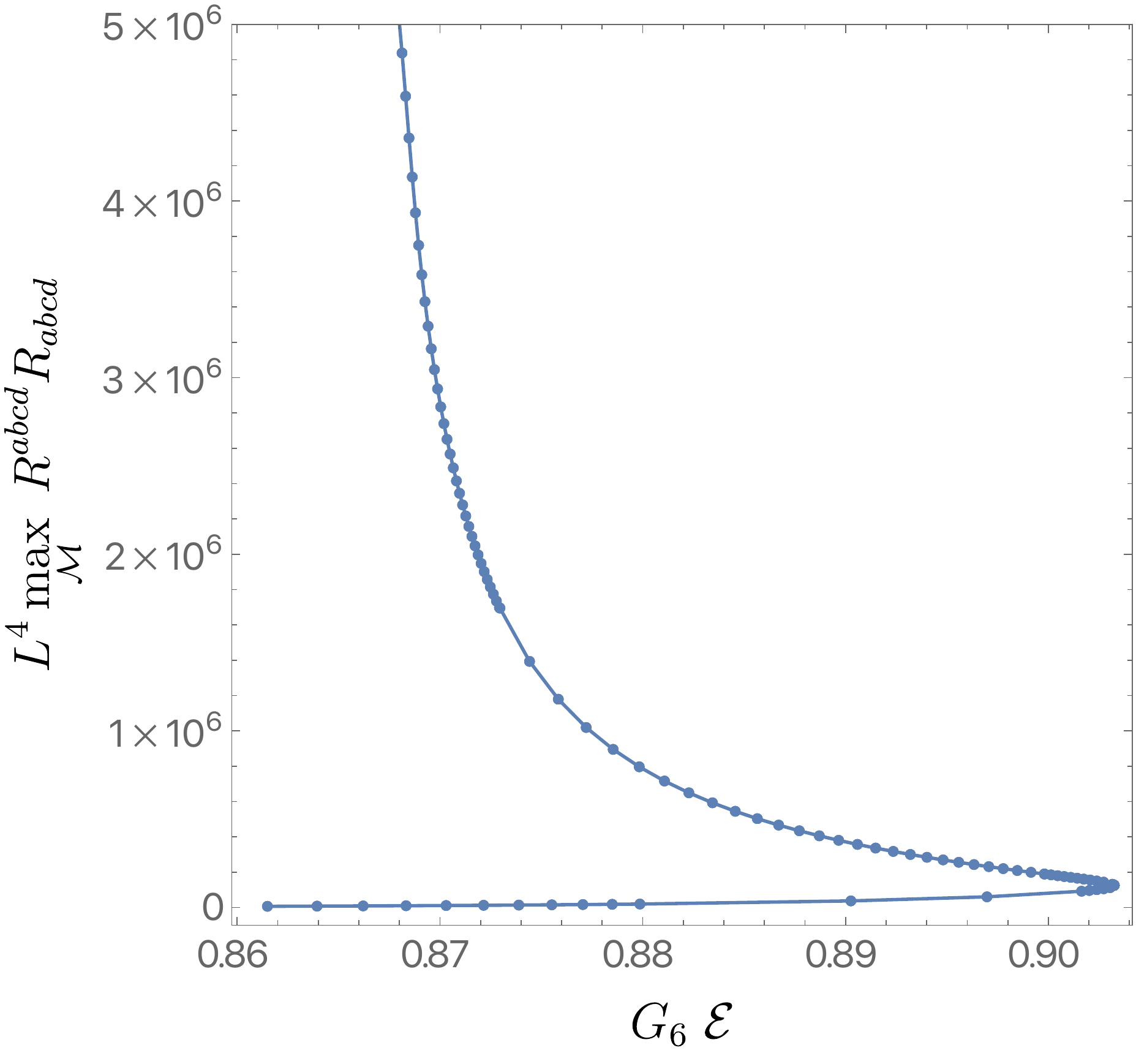}
\caption{Maximum value of the normalised Kretschmann scalar for black resonator strings as a function of $\mathcal{E}$ at fixed $\widetilde{\Omega}_H=0.5$. A turning point is clearly visible around $G_6\mathcal{E}\approx 0.903$.  The apparent blow-up of the Kretschmann scalar suggests that the phase space of black resonator strings ends in a curvature singularity.}
\label{fig:fold}
\end{figure}

There is a family of black resonator strings with maximum angular momenta.  These are given by the grey line in Fig.~\ref{Fig:phaseDiagResonator}.  At this line, there is a fold (or turning point).  The continued space of solutions now extends towards \emph{smaller} momenta, so the phase space just slightly below this gray turning-point curve has (at least) two black resonator string solutions.  Our results indicate that continuing the space of solutions further will result in a curvature singularity at finite $\Delta \mathcal{J}<\Delta \mathcal{J}_{\max}(\mathcal{E})$.  Evidence for this singularity are presented in Fig.~\ref{fig:fold} where we plot the maximum value of the normalised Kretschmann scalar as a function of $\mathcal{E}$ for fixed $\widetilde{a}=0.5$, where the aforementioned turning point can easily be identified.

The fact that the black resonator strings and the Kaluza-Klein geons are totally disconnected in the phase diagram deserves a comment. Typically, in superradiant black hole systems with black resonators (rotating or charged), the geons (or solitons) of the theory are the zero horizon radius limit (i.e. the zero entropy limit) of the resonator black holes \cite{Basu:2010uz,Dias:2011at,Dias:2011tj,Cardoso:2013pza,Dias:2015rxy,Dias:2016pma,Bhattacharyya:2010yg,Markeviciute:2016ivy,Dias:2022eyq}.  Sometimes, a resonator black hole can be well-approximated by a superposition of a `bald' black hole and a geon, so long as both components have the same angular velocity or chemical potential.  Evidently, such an approximate does not apply in the present Kaluza-Klein setup.

There are two sets of solutions which we have not discussed. First, there are rotating generalisations \cite{Kleihaus:2007dg} of the usual non-uniform strings of \cite{Wiseman:2002zc,Kleihaus:2006ee,Kalisch:2015via,Kalisch:2016fkm,Kalisch:2018efd}, which branch from the onset of the Gregory-Laflamme instability. The reason for not commenting on these solutions is that available results would suggest that these solutions have lower entropy than the corresponding Myers-Perry string at fixed values of $\mathcal{E}$ and $\mathcal{J}$. Second, we have not discussed Kaluza-Klein (a.k.a. localized) black holes \cite{Wiseman:2002ti,Sorkin:2003ka,Kudoh:2003ki,Kudoh:2004hs,Headrick:2009pv,Kalisch:2017bin}, which can also carry rotate. We leave the construction of these solutions, and their location in the respective phase diagram, to future work.
\section{Discussion and Conclusions  \label{sec:Discussion}}

In this paper, we considered vacuum Einstein gravity in 6-dimensions ($D=6$) and studied the phase diagram of asymptotically ${\cal M}^{1,4}\times S^1$ black string solutions with equal angular momenta along the two rotational planes. In this setup, Myers-Perry black strings are cohomogeneity-1, so it is the simplest model of a system that has both  Gregory-Laflamme and superradiant instabilities.
These instabilities were studied in detail in \cite{Dias:2022mde}, and the unstable regions were shown in Fig.~\ref{fig:stabilityDiag} and Fig.~\ref{Fig:zeroModeSuperGL}.

We constructed a new type of black string, black resonator strings, that branch from the onset of the $m=2$ superradiant instability.  These solutions have $SU(2)_L$ isometry, and a further isometry generated by the helical Killing vector field $K=\partial_t+\Omega_H \partial_\psi =\partial_\tau$.  That is, these black resonator strings are time periodic, and neither axisymmetric, nor translationally invariant along the $S^1$ direction.  We find that black resonator strings are entropically preferred over Myers-Perry black strings in regions of phase space where both solutions exist.

We also constructed Kaluza-Klein geons, which are horizonless, purely gravitational solutions that share the same symmetries as the black resonator strings.  Unlike other superradiant systems, the Kaluza-Klein geons are not the horizonless limit of black resonator strings.  The full phase diagram of black resonator strings, Kaluza-Klein geons, and Myers-Perry black strings were shown in Fig.~\ref{Fig:phaseDiagResonator}.

We now speculate on the time evolution of Myers-Perry black strings that are unstable to $m=2$ superradiant perturbations.  Even though black resonator strings have higher entropy than Myers-Perry black strings, it still remains unclear whether black resonator strings will serve as a final endpoint.  Some obstacles are (1) the existence of other more entropically preferred solutions, (2) the existence of further instabilities, (3) the status of radiative perturbations.

What other solutions could there be?  Conspicuously missing are the non-uniform strings and localised black holes that are associated to the Gregory-Laflamme instability. Previous studies of static black strings in $D=6$ dimensions suggest that the localised black holes often have the largest entropy, and that might remain true when rotation is accounted for. Also missing are $m\neq2$ black resonator strings, possibly from other sectors of perturbation theory.  Should any of these solutions have higher entropy that the Myers-Perry black string, then they would compete as a possible endpoint.  A further complication is that some of these solutions might be metastable, though not the entropically dominant.  The dynamics and ultimate endpoint would then depend heavily on the initial data.

Another question is whether there are competing instabilities.  This is certainly true for some regions of parameter space, as shown in Fig.~\ref{Fig:zeroModeSuperGL}.  Where Gregory-Laflamme instability and superradiant instabilities are both present, it is typical for the Gregory-Laflamme instability to dominate the early dynamics due to its higher growth rate.  Additionally, there are other superradiant instabilities, though it is notable in this case that there are only a finite number of them. This is unlike superradiant instabilities in some other systems, where there is an infinite number of unstable modes, often with arbitrarily high azimuthal mode number $m$ \cite{Dias:2011at,Cardoso:2013pza,Dias:2015rxy,Niehoff:2015oga,Chesler:2018txn,Chesler:2021ehz}.  For this reason, it seems likely that superradiant instabilities in this system will not cause a cascade to smaller and smaller length scales \cite{Dias:2015rxy,Niehoff:2015oga}.  The Gregory-Laflamme instability, on the other hand, can lead to a change in horizon topology, which necessitates a violation of weak cosmic censorship \cite{Horowitz:2001cz,Kleihaus:2007dg,Lehner:2010pn,Gubser:2001ac,Wiseman:2002zc,Sorkin:2004qq,Kudoh:2004hs,Sorkin:2006wp,Headrick:2009pv,Figueras:2012xj,Emparan:2015gva,Figueras:2022zkg}.

Yet another separate question is whether black resonator strings themselves are stable.  While they are certainly stable to the $m=2$ perturbations that generate them, there are already several other perturbations that cause instabilities within the Myers-Perry black string.  It seems likely that many of these instabilities will be present in black resonator strings as well.  Where and how these instabilities affect black resonator strings remains unclear.

A further complication is that many perturbations are radiative.  That is, perturbations can create gravitational waves that can reach null infinity, removing energy and angular momentum from the system.  The final state, therefore, does not necessarily need to have the same energy and angular momentum as the starting state, though the entropy of course must still increase.

Finally, we comment on whether our results apply to the $D=5$ Kerr black string.  In the Kerr case, we again have both a superradiant instability and the Gregory-Laflamme instability.  $D=5$ black resonator strings should likewise exist.  However, unlike the $D=6$ case, a single Kerr black string can be unstable to superradiant modes with arbitrarily high $m$.  This occurs because Kerr black holes, unlike Myers-Perry black holes, have bound orbits.  In this regard, the Kerr case is similar to other superradiant systems, and the evolution of that instability might lead to a cascade to smaller and smaller length scales.

\begin{acknowledgments}

O.J.C.D. acknowledges financial support from the  STFC ``Particle Physics Grants Panel (PPGP) 2018" Grant No.~ST/T000775/1. OD acknowledges the Isaac Newton Institute, Cambridge, and the organizers of its long term programme ``Applicable resurgent asymptotics: towards a universal theory" during which this work was completed. The work of T.I. was supported in part by JSPS KAKENHI Grant Number JP18H01214 and JP19K03871.
The work of K.M. was supported in part by JSPS KAKENHI Grant Nos. 20K03976, 21H05186 and 22H01217.
BW acknowledges support from ERC Advanced Grant GravBHs-692951 and MEC grant FPA2016-76005-C2-2-P. J.~E.~S. has been partially supported by STFC consolidated grant ST/T000694/1. The authors also acknowledge the use of the IRIDIS High Performance Computing Facility, and associated support services at the University of Southampton, in the completion of this work.
\end{acknowledgments}


\bibliography{refsResonator}{}
\bibliographystyle{JHEP}

\end{document}